\documentclass[acmsmall,natbib=true,screen=true]{acmart}

\AtBeginDocument{%
  \providecommand\BibTeX{{%
    \normalfont B\kern-0.5em{\scshape i\kern-0.25em b}\kern-0.8em\TeX}}}

\setcopyright{rightsretained}
\copyrightyear{2024}
\acmYear{2024}
\acmDOI{}
\acmJournal{TOIS}
\acmVolume{}
\acmNumber{}
\acmArticle{}
\acmMonth{7}

\usepackage{multirow}
\usepackage{multicol}
\usepackage{bbm}
\usepackage[inline]{enumitem} 
\usepackage{amsmath}
\usepackage{mathtools}
\usepackage{hyperref}       % hyperlinks
\usepackage{url}            % simple URL typesetting
\usepackage{csquotes}
\usepackage{natbib}
\usepackage{graphicx}
\usepackage{acronym}
\usepackage[skip=1pt]{caption}
\usepackage{cancel}
\usepackage{ulem}
\usepackage{xcolor}

\newcommand{\heading}[1]{\vspace*{1mm}\noindent\textbf{#1.}}

\acrodef{GR}{generative retrieval}
\acrodef{MLE}{maximum likelihood estimation}

\newcommand{\hcancel}[1]{%
  \ifmmode
    \text{\sout{\ensuremath{#1}}}%
  \else
    \sout{#1}%
  \fi
}
\renewcommand{\emph}[1]{\textit{#1}}

\author{Yubao Tang}
\orcid{0009-0003-8010-3404}
\email{tangyubao@ict.ac.cn}

\author{Ruqing Zhang}
\authornote{Research conducted when the author was at the University of Amsterdam.}
\orcid{0000-0003-4294-2541}
\email{zhangruqing@ict.ac.cn}  

\author{Jiafeng Guo}  
      \authornote{Jiafeng Guo is the corresponding author.}
\orcid{0000-0002-9509-8674}
\email{guojiafeng@ict.ac.cn
}    

\affiliation{
  \institution{
  Institute of Computing Technology, Chinese Academy of Sciences; 
  University of Chinese Academy of Sciences}
  \streetaddress{No. 6 Kexueyuan South Road, Haidian District}
  \city{Beijing}
  \country{China}
  \postcode{100190}
}

\author{Maarten de Rijke}
\orcid{0000-0002-1086-0202}
\email{m.derijke@uva.nl}

\affiliation{
  \institution{University of Amsterdam}
  \city{Amsterdam}
  \country{The Netherlands}
}

\author{Wei Chen}
\orcid{0000-0002-7438-5180}
\email{chenwei@ict.ac.cn
}

\author{Xueqi Cheng}
\orcid{0000-0002-5201-8195}
\email{cxq@ict.ac.cn 
}

\affiliation{
  \institution{
  Institute of Computing Technology, Chinese Academy of Sciences; 
  University of Chinese Academy of Sciences}
  \streetaddress{No. 6 Kexueyuan South Road, Haidian District}
  \city{Beijing}
  \country{China}
  \postcode{100190}
}

\renewcommand{\shortauthors}{Tang et al.}

\begin{document}

\title[Listwise Generative Retrieval Models via a Sequential Learning Process]{Listwise Generative Retrieval Models via a Sequential Learning Process}

\begin{abstract}
Recently, a novel \ac{GR} paradigm has been proposed, where a single sequence-to-sequence model is learned to directly generate a list of relevant document identifiers (docids) given a query. 
Existing \ac{GR} models commonly employ maximum likelihood estimation (MLE) for optimization: this involves maximizing the likelihood of a single relevant docid given an input query,  with the assumption that the likelihood for each docid is independent of the other docids in the list.  
We refer to these models as the pointwise approach in this paper. 
While the pointwise approach has been shown to be effective in the context of \ac{GR}, it is considered sub-optimal due to its disregard for the fundamental principle that ranking involves making predictions about lists. 
% it ignores the fact that ranking involves making predictions about lists of docids, rather than solely about individual docids. 
In this paper, we address this limitation by introducing an alternative listwise approach, which empowers the \ac{GR} model to optimize the relevance at the docid list level. 
% the difference between the ranked docid list produced by a \ac{GR} model and the ground-truth ranked list is highlighted.
Specifically,
we view the generation of a ranked docid list as a sequence learning process: at each step we learn a subset of parameters that maximizes the corresponding generation likelihood of the $i$-th docid given the (preceding) top $i-1$ docids. 
To formalize the sequence learning process, we design a positional conditional probability for \ac{GR}. 
To alleviate the potential impact of beam search on the generation quality during inference, we perform relevance calibration on the generation likelihood of model-generated docids according to relevance grades. 
We conduct extensive experiments on representative binary and multi-graded relevance datasets.
Our empirical results demonstrate that our method outperforms state-of-the-art \ac{GR} baselines in terms of retrieval performance.
\end{abstract}




\maketitle

\acresetall

\section{Introduction}

\iffalse
Along with the development of representation learning in IR, dense retrieval \cite{zhan2020repbert,luan2021sparse,bentley1975multidimensional,nie2020dc} has become a popular paradigm to improve retrieval performance. 
With learned representations of queries and documents, a document index can be constructed and a dot-product or cosine function is adopted to perform efficient similarity search in the representation space. 
Despite the effectiveness of dense retrieval, unresolved yet crucial challenges pertaining to its effectiveness and efficiency remain: 
\begin{enumerate}[label=(\roman*)]
    \item During training,  the usage of simplistic functions such as cosine similarity  overlooks the potential for fine-grained interactions between the query and the document, which may result in information loss and reduce retrieval accuracy. 
   
    \item During inference, they necessitate storing dense vectors for the whole corpus and matching a query against all the documents.  Unfortunately, this process comes with substantial computational costs and a significant memory footprint.
\end{enumerate}
\fi

\noindent%
Document retrieval plays a critical role in many information retrieval (IR) related tasks, e.g., web search \cite{msmarco,craswell2020overviewtrec2019} and question answering \cite{guu2020retrieval,sakai2011usingyahooqa}. 
It aims to return an initial set of potentially relevant documents from a large-scale document repository when given a query. 
Recently, a new retrieval paradigm called \acfi{GR} \cite{modelBased} for document retrieval has been proposed. 
The key idea is to fully parameterize different components of indexing and retrieval within a single consolidated model, in which the information of all the documents in a corpus is encoded into the model parameters. 
In essence, this paradigm formalizes the document retrieval task as a sequence-to-sequence (Seq2Seq) problem that directly maps string queries to relevant document identifiers (docids). 
Following the initial publication by \citet{modelBased}, many subsequent investigations \cite{DSI,zhuang2022bridgingdsiqg,chen2022corpusbrain,NCI,genre,seal} have showcased the potential of this novel paradigm. 
In comparison to traditional dense retrieval \cite{zhan2020repbert,luan2021sparse,bentley1975multidimensional,nie2020dc}, \ac{GR} has several advantages: 
\begin{enumerate}[label=(\roman*)]
    \item During training, such a consolidated model can be optimized directly in an end-to-end manner towards a global objective. 
    By generating docids token-by-token in an autoregressive fashion and conditioning them on the query, we can capture fine-grained interactions between the query and the document.

    \item During inference, the need for a complicated explicit index structure is eliminated. Instead, docid generation is performed using a vocabulary with tens of thousands of words, aligned with identifiers of all the documents in the corpus. 
    Such autoregressive decoding significantly reduces the memory space and computational costs. 
\end{enumerate}

\begin{figure}[t]
 \centering
 \includegraphics[width=0.8\textwidth]{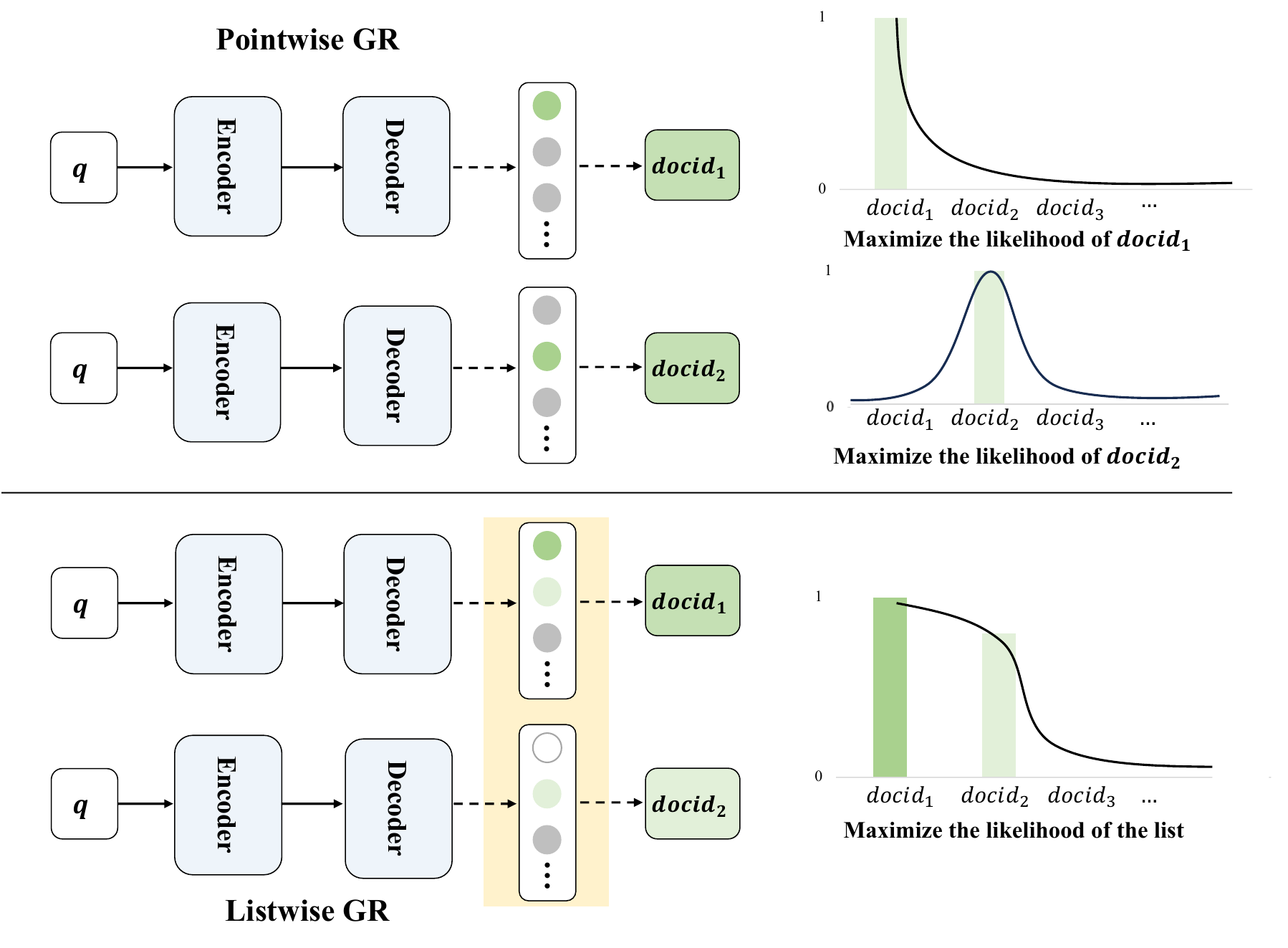}
  \vspace{-2mm}
 \caption{Optimization objectives. Assume that the following are given: a query $q$ and two ground-truth docids, $docid_1$ and $docid_2$, where $docid_1$ is more relevant than $docid_2$ to $q$.
Top: Most existing GR work relies on maximum likelihood estimation, by maximizing the likelihood of the target docid for each query-docid pair. All relevant docids $docid_1$ and $docid_2$ are treated equally, sharing similar likelihood values.
Bottom: A listwise objective (yellow rectangle) is designed for GR, directly modeling the ranked docid lists and incorporating positional information between $docid_1$ and $docid_2$ ($docid_1$ with darker green has a larger positional weight), resulting in a positional weighted likelihood.}
 \label{fig:contrast}
 \vspace{-5mm}
\end{figure}

The majority of existing \ac{GR} models relies on the standard Seq2Seq objective, i.e., maximum likelihood estimation (MLE) \cite{lamb2016professormle,lecun2015deepmle} with teacher forcing for learning. 
That is, during training, a number of queries are provided; each query is associated with a perfect ranked list of docids (in descending order of relevance scores); \ac{GR} models operate in a pointwise manner.
For example, as shown in Figure \ref{fig:contrast} (Top), existing works mainly focus on maximizing the likelihood of individual docids at a time. 
The final ranking is achieved by simply sorting the list based on the generated likelihood scores  of these docids. 
In essence, the score assigned to each docid is independent of the other docids for a given query. 
This approach suffers from several issues: 
First, the learning objective under the MLE criterion is formalized as minimizing errors in generation of docids, rather than minimizing errors in rankings of docids, making it inconsistent with  evaluation metrics like nDCG \cite{jarvelin2002cumulatedndcg}. 
Second, given a query, the assumption that the query-docid pairs are generated independently and identically distributed (i.i.d.) is a strong assumption. 
Thirdly, the number of query-docid pairs can vary greatly from one query to another, leading to a GR  model that is biased towards queries with a larger number of docid pairs \cite{10.1145/1273496.1273513listnet}. 

In this paper, we design a novel listwise approach to \ac{GR}, in which \emph{docid lists} instead of \emph{individual docids} are used as instances in learning, as shown in Figure \ref{fig:contrast} (Bottom).  
Inspired by listwise learning-to-rank  \cite{xia2008listwiselistmle,lan2014positionplistmle,10.1145/1273496.1273513listnet}, it is crucial to   effectively capture the difference between a ranked list of docids produced by a GR model and the ranked list given as the ground truth. 
To formalize the listwise loss function for \ac{GR}, our key idea is to view the problem of generating a ranked list of relevant docids as a sequential learning process: in each step we target to maximize the corresponding stepwise probability distribution. 
Specifically, at step 1, we aim to maximize the probability distribution that the top-1 docid is generated. 
At step $i>1$, we maximize the $i$-th probability distribution given the top $i-1$ docids. 
Leveraging the characteristics of GR, we define the probability distribution as the output sequence likelihood of generating each docid, token-by-token in an autoregressive fashion, and conditioned on the given query. 
To solve the sequential learning problem, we transform it into a single-objective optimization problem via linear scalarization, in which the position importance in ranking is highlighted \cite{lan2014positionplistmle}. 
By assigning appropriate weights to different ranking positions, the final listwise loss function can effectively emphasize the significance of each position and optimize the overall objective accordingly. 
The comparison between previous pointwise approaches and our proposed listwise approach for GR is illustrated in Figure~\ref{fig:contrast}. 
We refer to the GR model using the listwise loss function as ListGR. 

At inference time, the trained ListGR model uses beam search to generate a ranked list of potentially-relevant docids, which are based on possibly erroneous previous steps. 
However, in the proposed listwise loss function, the predictive probability of each reference docid is maximized given the gold sub-sequence before it. 
To solve this decoding inconsistency problem, we propose to perform relevance calibration to re-train the model with a relevance calibration objective. 
This objective aims to calibrate the likelihood of generated candidate docids to better align with ground-truth ranked lists according to their relevance grades to the query. 

Our main contributions are the following: 
\begin{enumerate}[label=(\roman*)]
    \item  To the best of our knowledge, this is the first proposal for a listwise approach specifically designed for GR. 
    \item  We formulate a listwise learning objective for GR,  by directly minimizing the expected loss defined on the predicted docid list and the ground-truth list, and taking into account position information. 
    \item Our experimental results on five representative retrieval datasets demonstrate the effectiveness of our method, particularly on datasets with multi-graded relevance. Compared to the current state-of-the-art pointwise GR method, NCI, our approach achieves a significant improvement of 15.8\% in terms of nDCG@5 on the ClueWeb 200K dataset.
\end{enumerate}

\noindent%
The remainder of the paper is structured as follows. Section~\ref{sec:preliminaries} introduces preliminary concepts necessary for understanding the proposed method. Section~\ref{sec:methods} outlines the details of our proposed method. Section~\ref{sec:experimental-settings} describes the experimental setup. Section~\ref{sec:experimental-results} presents the experimental results and analysis, highlighting the performance of our method compared to existing approaches. Section~\ref{sec:related-work} presents an overview of related work in the field. Finally, Section~\ref{sec:conclusion} provides a summary of the paper and  discusses limitations and potential future research directions.

\section{PRELIMINARIES}
\label{sec:preliminaries}

We first recall the basic idea of the GR paradigm and of listwise algorithms that have been widely adopted in learning-to-rank; Table \ref{tab:notations} lists the most important notations used in the paper.

\begin{table}[t]
% \small
    \caption{Important notation.}
    \label{tab:notations}
    \centering
    % \color{blue}
% \renewcommand{\arraystretch}{0.9}%row space 
   \setlength\tabcolsep{5pt}
    \begin{tabular}{cl}
    \toprule
    $D$ & Document set \\
    $d$ & Document in $D$ \\
    $I_D$ &Docid set corresponding to $D$\\
    $id$ &Docid in the docid set $I_D$\\
    $id^{(i)}$ & $i$-th ranked docid in a ranked docid list \\
    $\pi_q$ &Ground-truth docid list of $q$   \\
    $\pi_Q$ &Set of ground-truth docid lists of $Q$ \\
    $Q$ &Query set \\
    $q$ &Query\\
    $g_\theta$ &GR model parameters\\
    $w_t$ &Token in the docid $id$\\
    $x$ &Document set to be further ranked for a query \\
    $\pi_y$  & Ground-truth permutation of documents $x$ for a query \\
    $y_i$ &Ranking position of document $d_i$ in $\pi_y$ \\
    $y^{-1}(i)$ & Index identifier of documents in the $i$-th position of $\pi_y$ \\
    $h_\psi$ &Learning to rank function  \\
    
        \bottomrule
    \end{tabular}
\end{table}

\subsection{Generative retrieval}

Generative retrieval (GR) aims to directly generate a ranked list of docids for a given query using a text-to-text model. 
In the following, we summarize the model architecture, training, and inference process of GR.  

\subsubsection{\textbf{Model architecture.}} 
In existing approaches, the GR model, represented as $g_\theta$, usually makes use of a transformer-based encoder-decoder architecture to answer queries.  
The encoder is responsible for processing the input sequence, i.e., query or document, and extracting meaningful representations to capture the essential topics. 
Based on the representation produced by the encoder, the decoder is responsible for generating the target docid. 

\subsubsection{\textbf{Document identifiers (docids)}}
\citet{DSI} propose two primary document identifiers to represent documents:
\begin{enumerate}[label=(\roman*)]
\item Arbitrary unique integers without explicit semantic connections to the corresponding documents \cite{DSI}.
\item Structured semantic numbers that carry semantic associations with the documents, often obtained through techniques like hierarchical k-means clustering \citep{DSI, NCI}.
\end{enumerate}
Incorporating semantic associations between docids and documents improves the retrieval process \citep{DSI,NCI, genre, seal}. 
In this work, we adopt the structured semantic numbers for docid representation and we leave a detail discussion  of the docid generation process to Section~\ref{subsec:implementation-details}.
Recently, alternative forms of docids such as n-grams and titles have been proposed. A comprehensive explanation of these docids can be found in Section \ref{sec:related-work}.

\iffalse
\begin{table}[h]
% \small
    \caption{\textcolor{blue}{Important notations}}
    \label{tab:notations}
    \centering
    \color{blue}
% \renewcommand{\arraystretch}{0.9}%row space 
   \setlength\tabcolsep{8pt}
    \begin{tabular}{cl|cl}
    \toprule
    $D$ & Documnet set & $\pi_q$ &Ground-truth docid list of $q$\\
    $d$ & Document in $D$ & $\pi_Q$ &Set of ground-truth docid lists of $Q$\\
    $I_D$ &Docid set corresponding to $D$& $x$ &Document set to be further ranked for $q$\\
    $id$ &Docid in the docid set $I_D$& $X$ & Document set to be further ranked for $Q$\\
    \multirow{2}{*}{$id^{(i)}$}  & \multirow{2}{*}{\begin{tabular}{@{}c@{}}Docid ranked at the i-th position \\ in a ranked list\end{tabular}} & \multirow{2}{*}{$\pi_y$}  & \multirow{2}{*}{\begin{tabular}{@{}c@{}}Ground-truth permutation of documents \\for a query\end{tabular} } \\ 
    & & &\\
    $Q$ &Query set&$M$ &Set of relevance grades of documents \\
    $q$ &Query& $y$ &Position of a document in $\pi_y$\\
    $g_\theta$ &GR model parameters&  $Y$ &Set of document positions for $Q$\\
    $w_t$ &Token in the docid $id$& $H$ &Ranking function space\\
    $h_\psi$ &Ranking function &  \\

        \bottomrule
    \end{tabular}
\end{table}
\fi

\subsubsection{\textbf{Training and optimization}}\label{secsec:train-and-opt}

Maximum likelihood estimation (MLE) is widely employed in current GR methods to optimize two  main  tasks, i.e., the indexing task and the retrieval task, via maximizing the likelihood estimation of the target docid, given a document or query. 

\heading{Indexing task} To memorize the corpus, the GR model $g_\theta$  takes the document $d$ in the document set $D$ as the input, and outputs its corresponding docid $id$ in the docid set $I_D$ with MLE optimization algorithm, defined as, 
\begin{equation}\label{eq:indexing}
    \mathcal{L}_\mathit{Indexing}(D,I_D;g_\theta) = - \sum_{d \in D} \log P(id\mid d;g_\theta),
\end{equation}
where $P(id\mid d;g_\theta)$ is the likelihood of generation docid $id$, 
\begin{equation}\label{eq:teacher-forcing-indexing}
    P(id\mid d;g_\theta) = \prod_{t \in [1,|id|]} P(w_t \mid d,w_{<t};g_\theta),
\end{equation}
where $w_t$ is the $i$-th ground-truth token in the $id$, and $w_{<t}$ represents the tokens before the $i$-th one in the $id$.

\heading{Retrieval task} A query $q$ in the query set $Q$ can have one or multiple associated docids, and these docids may possess varying degrees of relevance. 
For $q$, it has a ground-truth docid list, $\pi_q=[id^{(1)},id^{(2)},\ldots]$, in descending order of relevance, where $id^{(1)}$ is the docid ranked at the first position, and $id^{(2)}$ is the docid ranked
at the second position, and so on. 
We denote the set of relevant docids for all the queries $Q$ as $\pi_Q$.
Relevance grades for documents are non-negative integers.
A relevance grade of $0$ indicates that the document is irrelevant to the query. 
The higher the integer value, the greater the relevance of the document to the given query. 
And $M(d)$ denotes the relevance grade of the document $d$ to a query. 
To achieve the retrieval task effectively, the GR model also leverages MLE to learn how to map the query $q$ in the query set $Q$ to relevant docids, defined as, 
\begin{equation}\label{eq:retrieval}
    \mathcal{L}_\mathit{Retrieval}(Q,\pi_Q;g_\theta) = - \sum_{q\in Q,id \in \pi_q}
     \log P(id\mid q;g_\theta),
\end{equation}
where $P(id\mid q;g_\theta)$ is similar to Eq.~\eqref{eq:teacher-forcing-indexing}, defined as
% with the difference being that the input is the query $q$ instead of the document.
%
\begin{equation}\label{eq:teacher-forcing-retrieval}
    P(id\mid q;g_\theta) = \prod_{t \in [1,|id|]} P(w_t \mid q,w_{<t};g_\theta). 
\end{equation}
Finally, the total loss incurred during training a GR model is a combination of the indexing loss and the retrieval loss, i.e., 
\begin{equation}\label{eq:indexing-retrieval}
    \mathcal{L}_{\mathit{Total}}(Q,D,I_{D}) =  \mathcal{L}_\mathit{Indexing}(D,I_D;g_\theta)+\mathcal{L}_\mathit{Retrieval}(Q,\pi_Q;g_\theta).
\end{equation}

\subsubsection{\textbf{Inference}}\label{secsec:inference}
During inference, given a query, the GR model usually uses beam search \citep{2020Beam} to generate the top-$n$ ranked docids in an autoregressive manner, in descending order based on the conditional probability of each output. 
Note that, when generating the next token, the model relies on the former generated token, rather than the ground-truth token. 

\subsubsection{\textbf{Discussion}} In current GR methods, MLE is primarily used to train query-docid pairs (as shown in Eq.~\eqref{eq:retrieval}), which is a pointwise approach. 
This approach, however, is limited in its ability to support the model in generating the single most relevant docid even when a query has multiple relevant docids. 
During inference, the goal of the retrieval task is to obtain a ranked docid list, where the docids are ordered based on their relevance to the query. 
The pointwise approach fails to guarantee an optimal ordering of docids within the list. 

To address this limitation and enhance the capability of the GR model to generate a high-quality ranked docid list, this work focuses on modeling and optimizing the relevance at the list level. 
By shifting the optimization objective from a pointwise perspective to a listwise perspective, we aim to further improve the overall effectiveness of the GR models.

\subsection{ListMLE algorithm}

In learning-to-rank (LTR), listwise approaches emphasize optimizing the entire ranked list of items for overall ranking performance. Listwise approaches recognize that the order in which items are presented in the list is crucial for accurate ranking. 
In the following, we describe a related algorithm for our work, including listMLE and position-aware listMLE. 

\subsubsection{\textbf{ListMLE}} 

Formally, suppose $x=\{d_1, \ldots, d_n\} \in X $ is the subset of corpus $D$ to be further ranked, obtained from an initial document retrieval step. 
And $\pi_y=[y_1, \ldots,y_n] \in Y$ is the corresponding ground-truth permutation of these documents, where $y_i$ is the position of $d_i$, and $y^{-1}(i)$ is the index identifier of documents in the $i$-th position of $\pi_y$. 
Listwise LTR aims to learn a ranking function $h_\psi:X \rightarrow Y$, where $\psi$ are the function parameters and $H$ is the corresponding function space (i.e., $h \in H$), that can minimize the expected risk. 

ListMLE \cite{xia2008listwiselistmle} is a widely-used framework for listwise ranking that introduces a parameterized exponential probability distribution over all possible permutations, given the ranking function $h_{\psi}$. And it leverages negative log likelihood of the ground truth list as the loss function, defined as:
\begin{equation}
    \mathcal{L}(x,\pi_y;h_{\psi})= -\log P(\pi_y\mid x;h_{\psi}).
\end{equation}
According to the Plackett-Luce model \cite{plackett1975analysisplackett,luce2012individualluce}, which is a distribution over permutations $\pi_y$,
$P(\pi_y\mid x;h_{\psi})$ can be defined as:
\begin{equation}
    P(\pi_y\mid x;h_{\psi}) = \prod^n_{i=1} \frac{\exp (h_{\psi}(x_{y^{-1}(i)}))}{\sum^n_{k=i} \exp(h_{\psi}(x_{y^{-1}(k)}))}.
    \label{eq:Plackett-Luce}
\end{equation}
The probability of a list can be deconstructed into the product of stepwise conditional probabilities. 
Each $i$-th conditional probability represents the likelihood of 
% the document being ranked at the $i$-th position, given that the preceding $i-1$ documents are ranked correctly, 
a document being ranked at the $i$-th position, given that the preceding documents are ranked appropriately up to that point.
i.e.,
\begin{align}\label{eq:listmle}
    P(\pi_y\mid x;h_{\psi}) &= P(y^{-1}(1),\ldots, y^{-1}(n)\mid x;h_{\psi}) \\
    &=P(y^{-1}(1) \mid x;h_{\psi})\prod^n_{i=2}P(y^{-1}(i)\mid x,y^{-1}(1),\ldots,y^{-1}(i-1);h_{\psi}),
\end{align}
where
\begin{align}
    P(y^{-1}(1)\mid x;h_{\psi}) &=  \frac{\exp (h_{\psi}(x_{y^{-1}(1)}))}{\sum^n_{k=1} \exp (h_{\psi}(x_{y^{-1}(k)}))}, 
    \\
    P(y^{-1}(i)\mid x,y^{-1}(1),y^{-1}(2),\ldots,y^{-1}(i-1);h_{\psi}) &= \frac{\exp(h_{\psi}(x_{y^{-1}(i)}))}{\sum^n_{k=i} \exp (h_{\psi}(x_{y^{-1}(k)}))}, \forall i=2,\dots,n.
\end{align}

\subsubsection{\textbf{Position-aware ListMLE}} 
ListMLE, despite its effectiveness, ignores the significance of position importance~\cite{lan2014positionplistmle}. 
Recognizing the impact of item positions for ranking, an advanced listwise ranking algorithm called \emph{position-aware ListMLE} \cite[p-ListMLE,][]{lan2014positionplistmle} has been developed to take into account position information. 
p-ListMLE considers the ranking process as a sequential procedure: it operates by maximizing the probability of correctly ranking the top 1 document with a weight assigned to the top position. 
Subsequently, it focuses on maximizing the probability of correctly ranking the $i$-th document, considering the corresponding position weight, assuming that the top $i - 1$ documents have been ranked correctly.  
This loss function of the process is formally defined as:
\begin{equation}
\label{eq:plistmle}
    \begin{split}
        \mathcal{L}_p(x,\pi_y;h_{\psi}) ={} & -\alpha(1)\log P(y^{-1}(1)\mid x;h_{\psi}) - {}\\
        & \sum^n_{i=2}\alpha(i)\log P\left(y^{-1}(i)\mid x, y^{-1}(1), \ldots, y^{-1}(i-1);h_{\psi}\right),
    \end{split}    
\end{equation}
where $\alpha(\cdot)$ is a decreasing function, i.e., $\alpha(i)>\alpha(i+1)$. 

To ensure consistency with the target metric, such as normalized discounted cumulative gain (NDCG),  $\alpha(\cdot)$ is defined as the gain function $\alpha(i) = Gain(i) = 2^{n-i}-1$, which assigns larger weights to documents with higher relevance grades. 
Combining the Plackett-Luce model~\eqref{eq:Plackett-Luce} with the above loss \eqref{eq:plistmle}, the optimization objective is to minimize the following likelihood loss function: 
\begin{equation}
    \mathcal{L}_p(x,\pi_y;h_{\psi}) = \sum^n_{i=1} \alpha(i)\left(-h_{\psi}(x_{y^{-1}(i)})+ \log\left(\sum^n_{k=i} \exp(h_{\psi}(x_{y^{-1}(k)}))\right)\right).
\end{equation}

% \subsection{Document Retrieval}
% Assume $D$ is a large-scale corpus and $d$ is an individual document. Meanwhile $Q$ is a query set, and $q$ is an individual query.
% Given a query, the document retrieval task aims to return top-$n$ ranked documents, in descending order of relevance score, i.e.,
% \begin{equation}
%     \pi_{f}(q) := [d^{(1)},\cdots,d^{(n)}] = [{\arg \max}_{d}^{(1)}f_{\phi}(q,d), \dots, {\arg \max}_{d}^{(n)}f_{\phi}(q,d),],
% \end{equation}
% where ${\arg \max}_{d}^{(i)}f_{\phi}(q,d) $ denotes the $i$-th $(1\leq i\leq n)$ ranked document $d^{(i)}$ for a query $q$ over $D$ given by the retriever $f$ with its parameters $\phi$. 
% During training, retrieval models leverage observed query-document pairs as the supervised signals to train the model.  For a query $q$, it may contains multiple relevant documents, denoted as $D_q$. And we denote all the relevant documents for the query set $Q$ as $D_Q$.
% For a query, its relevant documents may own different relevance grades. We denote the relevance grades set as $M=[0,1,2,\dots]$, and $M(d)$ denotes the relevance label of $d$. 

% For dense retrieval methods, $f_{\phi}$ usually leverages a dual-encoder architecture, and needs to learn a query and document encoder with observed query-document pairs. 
% During inference, the query and document encoder map the query and document into embeddings, respectively. Further, the similarity of embeddings of the query and documents is calculated, i.e., dot product and cosine similarity, as the relevance score. 

\section{Our approach}
\label{sec:methods}

In this section, we present novel listwise generative retrieval models via a sequential learning process. 
We first provide an overview of our method and then describe the training and re-training stages in detail. 

\subsection{Overview}

\begin{figure}[t]
 \centering
 \includegraphics[width=0.8\textwidth]{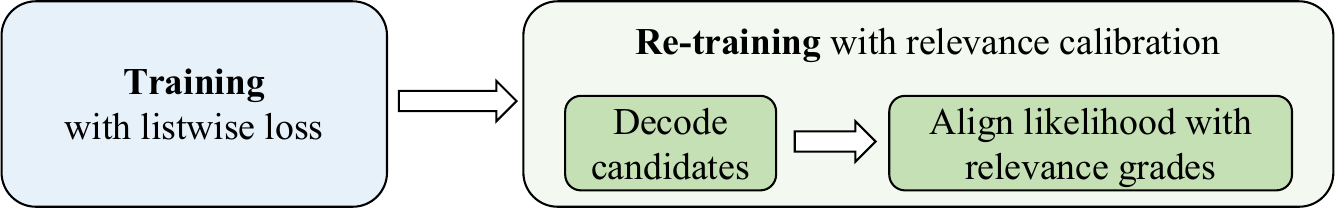}

 \caption{Overview of the two-stage listwise learning methods, which consists of a training stage using listwise loss and a re-training stage with relevance calibration based on the trained model.}
 \label{fig:two-stage}
\end{figure}

In this paper, we propose a \emph{listwise GR approach} (ListGR for short), in which docid lists instead of individual docids are used as instances in learning. 
ListGR includes a two-stage optimization process, i.e., training with position-aware ListMLE and re-training with relevance calibration. 
The overall optimization process is illustrated in Figure \ref{fig:two-stage}. 

To accurately represent listwise relevance, we first establish the position-aware conditional probability of a docid ranked at a particular position with respect to a given query, and employ position-aware ListMLE \cite{lan2014positionplistmle} to train the GR model. 
To address the decoding inconsistency between the proposed listwise loss function and the beam search decoding, we further retrain the model with relevance calibration techniques for a generated docid list.

\subsection{Training with listwise loss function}

Inspired by listwise LTR algorithms \cite{xia2008listwiselistmle,lan2014positionplistmle,10.1145/1273496.1273513listnet}, our key idea is to view the docid ranking problem as a sequential learning process, with each step targeting to maximize the corresponding stepwise probability distribution. 
In the following, we firstly define the conditional probability distribution of each ground-truth docid with each step, and then apply it to model the ranking list. 

\subsubsection{\textbf{Positional conditional probability}}
Given a query and its ground-truth docid list \(\pi_q\), for each docid \(id^{(i)}\) in the list, we first obtain the estimated log-probability of generating \(id^{(i)}\) for the given query (based on Eq.~\eqref{eq:teacher-forcing-retrieval}), regardless of the position of \(id^{(i)}\), and perform length normalization, denoted as,
\begin{equation}\label{eq:log-prob-gen-docid}
\tilde{P}(id^{(i)}\mid q;g_{\theta}) = \frac{\log \prod_{t \in [1,|id^{(i)}|]} P(w_t \mid q,w_{<t};g_\theta)}{|id^{(i)}|},
\end{equation}
where $w_t$ is the $i$-th ground-truth token in the $id^{(i)}$, $w_{<t}$
 represents the tokens before the $i$-th one in the $id^{(i)}$.
Based on Eq.~\eqref{eq:log-prob-gen-docid}, we further define the positional likelihood, that is, the probability of generation of $id^{(i)}$ being ranked at position $i$. 
More specifically, it is the conditional probability distribution of the GR model generating the ground-truth docids from the $i$-th to the $n$-th, conditioned on the query.
It also represents that the preceding $i-1$ docids are generated at the right positions. 
The sequential learning process for docid  ranking can be summarized as follows: 

\begin{description}
    \item[\textbf{Step 1}:] Maximizing the following top-1 positional conditional probability: 
    \begin{equation}\label{eq:listGR-step-1}
    P(id^{(1)}\mid q;g_{\theta}) = \frac{\exp( \tilde{P}(id^{(1)}\mid q;g_{\theta}))}{\sum^n_{j=1} \exp(\tilde{P}(id^{(j)}\mid q;g_{\theta}))}.
    \end{equation}
    Please note that Eq.~\eqref{eq:log-prob-gen-docid} only considers the generation of $id^{(i)}$ conditioned on the query without considering its ranking position in the list , while Eq.~\eqref{eq:listGR-step-1} requires $id^{(i)}$ to be ranked at the $i$-th position.

    \item[\textbf{Step $\boldmath{i}$}:]  For $ i = 2,\dots,n$, we maximize the following $i$-th positional conditional probability, 
    \begin{equation}\label{eq:listGR-step-i}
    P(id^{(i)}\mid q,id^{(1)},\dots,id^{(i-1)};g_{\theta}) = \frac{\exp( \tilde{P}(id^{(i)}\mid q;g_{\theta}))}{\sum^n_{j=i} \exp(\tilde{P}(id^{(j)}\mid q;g_{\theta}))}. 
    \end{equation}
    The learning process ends at step $n+1$. 
\end{description}

\subsubsection{\textbf{Listwise probability with position importance}}

To transform the above sequential optimization problem into a single-objective optimization problem, we define the likelihood of the ground-truth docid list $\pi_q$ for a query.  
The likelihood of generating $\pi_q$ is defined as the product of positional conditional probabilities of different docids. 
Higher positions are more important and, therefore, we assign the corresponding positional conditional probabilities with higher weights. 
Therefore, for a query $q$, the optimization problem is to minimize the probability of generating $\pi_q$ with negative log likelihood as follows:
\begin{equation}\label{eq:listwise-GR}
\begin{split}
 \min_{g_\theta} &-\log P(\pi_q\mid q;g_{\theta}) \\
   &= -\alpha(1)\log P(id^{(1)}\mid q;g_\theta) -\sum^n_{i=2} \alpha(i) \log P\left(id^{(i)}\mid q,id^{(1)},\dots,id^{(i-1)};g_{\theta}\right), 
\end{split}
\end{equation}
where the weight $\alpha(\cdot)$ is a decreasing function; following \cite{lan2014positionplistmle}, we set $\alpha(i) = 2^{n-i}-1$. 
Incorporating the probability based on Plackett-Luce model as described in Eq.~\eqref{eq:listGR-step-1} and~\ref{eq:listGR-step-i} into the above optimization problem, we obtain the final listwise loss function:
\begin{equation}\label{eq:listGR-loss}
    \mathcal{L}_\mathit{List}(q,\pi_q;g_{\theta})= \sum^n_{i=1} \alpha(i)\left(-\tilde{P}(id^{(i)}\mid q;g_{\theta})+ \log\left(\sum^n_{k=i} \exp(\tilde{P}(id^{(k)}\mid q;g_{\theta}))\right)\right).
\end{equation}
The total loss function of a query set $Q$ is $\mathcal{L}_\mathit{List}(Q,\pi_Q;g_\theta)=\sum_{q\in Q} \mathcal{L}_\mathit{List}(q,\pi_q;g_{\theta})$.

\heading{Discussions}
For retrieval tasks, our listwise approach $\mathcal{L}_\mathit{List}$ is different from the pointwise approach used in existing GR works. 
\begin{enumerate*}[label=(\roman*)]
\item The existing pointwise approach aims to maximize the likelihood probability of generating relevant docids with MLE for a query. Simultaneously, it suppresses the probability of other irrelevant tokens. When dealing with multi-graded relevance datasets, this method treats docids of different relevance grades as equally important.
\item In contrast, our listwise approach, $\mathcal{L}_\mathit{List}$ loss (Eq.~\eqref{eq:listGR-loss}) maximizes the likelihood probability of the ground-truth docid list. Additionally, it assigns corresponding positional weights to different docids based on their relevance grades using p-ListMLE. This enables the model to have better discriminative ability for fine-grained relevance grades.
\end{enumerate*}
Therefore, our approach aligns more closely with the goal of GR, which is to generate a relevant docid list given a query.

\subsubsection{\textbf{Training loss}}
To model the aforementioned listwise relevance, the GR model also needs to learn the fundamental indexing task with the loss defined in Eq.~\eqref{eq:indexing} and the retrieval task with the loss defined in Eq.~\eqref{eq:retrieval}. 
Taken together, the total loss for the training stage is defined as:
\begin{equation}\label{eq:training}
    \mathcal{L}_\mathit{Training}(Q,D,I_D,\pi_Q;g_\theta) = \mathcal{L}_\mathit{List}(Q,\pi_Q;g_{\theta}) +  \mathcal{L}_\mathit{Indexing}(D,I_D;g_\theta) + \mathcal{L}_\mathit{Retrieval}(Q,\pi_Q;g_\theta).
\end{equation}

\subsection{Re-training with relevance calibration}
After training with a listwise loss function, the GR model gains a better discriminative ability for ranked lists of docids than the previous pointwise approach. 
However, a decoding inconsistency problem arises \cite{bengio2015scheduledexpro}.
During training, the proposed listwise loss leverages the preceding ground-truth tokens to  generate the subsequent token.
During inference, the model relies solely on the preceding generated tokens without access to ground-truth tokens.
This decoding inconsistency may result in the generated list not being ideal in terms of its ranking according to relevance. 
Besides, larger beam sizes would cause shorter lengths and worse generation quality \cite{yang2018breaking,zhao2022calibrating}. 

To further improve the quality of the ranked list, we propose to calibrate the generated list, in which the key idea is to align candidates' likelihoods according to their relevance grades to the query. 
Specifically, we utilize the model trained with Eq.~\eqref{eq:training} for re-retraining, denoted as $\widehat{g}_\theta$. 
And for a given query $q \in Q$, a ranked docid list is generated with the beam search strategy, denoted as $\widehat{\pi}_q=[\widehat{id}^{(1)},\dots,\widehat{id}^{(n)}]$. 
We perform both token-level calibration and sequence-level relevance calibration as follows. 

\subsubsection{\textbf{Token-level relevance calibration.}}
For correctly predicted docids, tokens within docids with higher relevance grades are assigned with higher likelihood weights. 
For incorrectly predicted docids, the generation probability of their tokens should approach zero. 
Formally, we define the token-level relevance calibration loss as,
\begin{equation}
    \mathcal{L}_\mathit{Token}(Q,\widehat{\pi}_{Q};\widehat{g}_\theta) = -\sum_{q\in Q}\sum_{\widehat{id}\in \widehat{\pi}_{q}} \sum_{w_t \in \widehat{id}}P_\mathit{true}(w_t \mid q, w_{<t})\log P(w_t \mid q, w_{<t};\widehat{g}_\theta),
\end{equation}
where $w_t$ is the $t$-th generated token in $\widehat{id}$, $w_{<t}$ represents tokens before the $t$-th token, and $\widehat{\pi}_{Q}$ is the generated docid list for all queries in $Q$. 
Moreover, $P_{true}(w_t \mid q, w_{<t})$ is the weight of generating token $w_t$ computed as follows, given two different candidate docids in $\pi_q$, $\widehat{id}^{(i)}$ and  $\widehat{id}^{(j)}$:
\begin{equation}
\label{eq:token-level-cases}
  \left\{
    \begin{array}{ll}
        P_\mathit{true}(w_t \mid q, w_{<t}) = 1-\frac{1}{\left(M\left(\widehat{id}^{(i)}\right)+1\right)^2}, &\forall w_t \in \widehat{id}^{(i)}, \text{ if }\widehat{id}^{(i)} \in \pi_q \\
        \sum_{\widehat{id}^{(i)} \in \widehat{\pi}_{q}} P_\mathit{true}(w_t \mid q, w_{<t}) =  \beta, & \forall w_t \in \widehat{id}^{(i)}, \text{ if }\widehat{id}^{(i)} \notin \pi_q \\
        P_\mathit{true}(w_t \mid q, w_{<t}) >  P_\mathit{true}(w_m \mid q, w_{<m}), 
       &\forall w_t \in \widehat{id}^{(i)} \in \widehat{\pi}_q, w_m \in \widehat{id}^{(j)} \in\widehat{\pi}_q, \\
       &\text{if }M\left(\widehat{id}^{(i)}\right) > M\left(\widehat{id}^{(j)}\right) \\
    \end{array}
    \right.
\end{equation}
where $\beta$ is a hyperparameter close to zero, and $M\left(\widehat{id}^{(i)}\right)$ is the relevance grade of $\widehat{id}^{(i)}$, defined in Section \ref{secsec:train-and-opt}. Additionally, $w_m$ represents the $m$-th token of $\widehat{id}^{(j)}$, and $w_{<m}$ represents tokens before the $m$-th token in $\widehat{id}^{(j)}$. The effect of each condition of this equation is as follows,
\begin{enumerate}[label=(\roman*)]
    \item For the first condition, if the generated $\widehat{id}^{(i)}$ is in the ground-truth ranking list $\pi_q$, we assign a higher weight $P_{true}$ to this docid, to support its generation. Specifically, this weight is a value less than 1, directly proportional to the relevance grade of the ground-truth label. The higher the relevance grade of the docid, the closer this weight is to 1. 
    
    \item For the second condition, if the predicted $\widehat{id}^{(i)}$ does not belong to the ground truth docid list, we assign a small weight to suppress its generation. Specifically, this weight $\beta$ is a small value less than 1, approaching 0.
    
    \item For the third condition, for any two docids in the candidate docid list, $\widehat{id}^{(i)}$ and $\widehat{id}^{(j)}$, both belonging to the ground truth ranking list, we adjust their relative weights based on their ground-truth relevance grades.  If the relevance grade of $\widehat{id}^{(i)}$ is higher than that of $\widehat{id}^{(j)}$, then the tokens of $\widehat{id}^{(i)}$ should have higher weights (i.e., $P_{true}(w_t\mid q,w_{<t})$) compared to weights (i.e., $P_{true}(w_m\mid q,w_{<m})$) of $\widehat{id}^{(j)}$.
\end{enumerate}

\subsubsection{\textbf{Sequence-level relevance calibration.}}
Differences in generation probabilities among distinct docids should correspond to differences in their relevance grades. 
Docids with higher relevance grades should be prioritized, resulting in a higher likelihood of being ranked higher and generated. 
Therefore, the sequence-level relevance calibration loss is, 
\begin{equation}\label{eq:retrain-seq}
    \mathcal{L}_\mathit{Seq} (Q,\widehat{\pi}_{Q};\widehat{g}_\theta) =\sum_i \sum_{j>i} \max\left(0,\widehat{g}_\theta (\widehat{id}^{(j)})-\widehat{g}_\theta (\widehat{id}^{(i)})+\lambda_{ij}\right),
\end{equation}
where 
$\widehat{g}_\theta (\widehat{id}^{(i)})$ is $ P(\widehat{id}^{(i)} \mid q;\widehat{g}_\theta)$ normalized by docid length, i.e., 
\begin{equation}
    \widehat{g}_\theta (\widehat{id}^{(i)})=\frac{\log P(\widehat{id}^{(i)} \mid q;\widehat{g}_\theta)}{|\widehat{id}^{(i)}|^\alpha}, 
\end{equation}
where $\alpha$ is the length penalty hyperparameter, 
$\forall i,j, 1<i<j\leq n$, and $\lambda_{ij}$ is the margin multiplied by the difference in rank position between the docids, i.e., $\lambda_{ij}=(j-i)\lambda$. 

\subsubsection{\textbf{Re-training loss}}
The final loss of the relevance calibration is defined as:
\begin{equation}\label{eq:retrain}
    \mathcal{L}_\mathit{Re\textit{-}training} (Q,\widehat{\pi}_Q;\widehat{g}_\theta) = \mathcal{L}_\mathit{Token}(Q,\widehat{\pi}_Q;\widehat{g}_\theta) + \gamma\mathcal{L}_{Seq} (Q,\widehat{\pi}_Q;\widehat{g}_\theta),
\end{equation}
where $\gamma$ is the hyperparameter of balancing the two losses.

In summary, our model is first trained using the listwise loss in Eq.~\eqref{eq:training} and then used to decode ranked docid lists for training queries. After re-training the model using the loss in Eq.~\eqref{eq:retrain}, inference is performed according to the approach described in Section \ref{secsec:inference}. 

\heading{Adaption to binary relevance data}
For binary relevance data, since the relevant docids for a query have the same relevance grade, a query may have one or multiple ground-truth docid lists, each containing only one relevant docid, i.e., the top-1 docid. 
Therefore, in the first training stage, the corresponding position weight $\alpha(1)$ in the listwise loss (Eq.~\eqref{eq:listGR-loss}) is set to 0 ($\alpha(i)=2^{(n-i)}-1$). Consequently, Eq.~\eqref{eq:training} reduces into Eq.~\eqref{eq:indexing-retrieval}. For binary relevance data, this is acceptable since it only contains docids with the same relevance grade.
Our main improvement for the binary relevance data is the relevance calibration Eq.~\eqref{eq:retrain} in the re-training stage. It could further optimize the generated candidate docid list, according to docids' relevance grade to the query. 
In future work, we will explore designing alternative weight functions to enable list-level enhancement for binary relevance data in the first stage as well.

\section{Experimental Settings}
\label{sec:experimental-settings}

In this section, we present the experimental settings, including datasets, baselines, model variants, evaluation metrics, and implementation details.

\subsection{Datasets} 

We utilize five widely-used ad-hoc retrieval datasets: 
\begin{enumerate}[label=(\roman*)]
    \item \textbf{ClueWeb09-B} (ClueWeb) \cite{clarke2009overviewclueweb} is a large-scale  web collection containing over 50 million documents. The topics are gathered from the TREC Web Tracks conducted from 2009 to 2011.
    \item \textbf{Gov2} \cite{clarke2004overviewGov} consists of approximately 150 queries and 25 million documents collected from the .gov domain web pages. The topics are accumulated from the TREC Terabyte Tracks from 2004 to 2006.
    \item \textbf{Robust04} \cite{voorhees2004overviewrobust} comprises 250 queries and 0.5 million news articles. The topics of the queries are collected from the TREC 2004 Robust Track.
    \item \textbf{MS MARCO Document Ranking} (MS MARCO) \cite{msmarco} is a comprehensive benchmark dataset for web document retrieval.
    \item \textbf{Natural Questions} (NQ) \cite{naturalquestion} includes natural language questions as queries and Wikipedia articles as documents. Following previous GR  studies \cite{DSI,seal,NCI}, we perform experiments on the NQ320K version of the dataset, containing 307,000 query-document pairs.  
\end{enumerate}

\heading{Dataset preprocessing}
For multi-graded relevance datasets, i.e., ClueWeb, Gov2, and Robust04 datasets, they are annotated with multi-graded relevance labels, indicating varying degrees of matching with the query intent or information need. 
Akin to \cite{DSI}, we sample subsets of the original ClueWeb, Gov2, and Robust04 corpora, each of size 200K, for our subsequent experiments. These sampled subsets are referred to as \textbf{ClueWeb 200K}, \textbf{Gov 200K}, and \textbf{Robust 200K}, respectively. The sampling process involves selecting annotated documents first and then randomly choosing additional documents from the remaining corpus, resulting in a total of 200K documents.

For binary relevance datsets, i.e., MS MARCO and NQ datasets, they have documents labeled with binary relevance, indicating whether a document is relevant or irrelevant to a query. 
For the MS MARCO dataset, following \cite{zhuang2022bridgingdsiqg,chen2023understanding}, we sample a sub-dataset, \textbf{MS MARCO 100K}, consisting of 100K documents, 97K training queries and 3K queries for testing. We sample the training and testing queries from the original training set and development set, respectively. For the NQ320K dataset, following \cite{NCI}, we utilize its open-source preprocessing code.\footnote{\url{https://github.com/solidsea98/Neural-Corpus-Indexer-NCI/blob/main/Data_process/NQ_dataset/NQ_dataset_Process.ipynb}} It removes special characters from the documents and performs cleaning and concatenation based on the document structure, such as titles, abstracts, and body text.
Table \ref{tab:datasets} provides statistics of the datasets used in experiments. 
%For ClueWeb, Gov2, and Robust04 datasets, we sample 0.2 million documents from the entire corpus due to their large size.

\begin{table}[h]
% \small
    \caption{Data statistics. 
    \#Grades denotes the number of relevance grades, e.g., highly relevant and relevant.
    \#Avg denotes the average number of multi-graded relevant documents for queries. 
    }
    \label{tab:datasets}
    \centering
   \setlength\tabcolsep{5pt}
    \begin{tabular}{llcccc}
         \toprule
       \textbf{Relevance Type} & \textbf{Dataset}  & \textbf{\#Queries} & \textbf{\#Documents} & \textbf{\#Grades} & \textbf{\#Avg}\\
        \midrule
      Multi-graded & Robust 200K  & 250 & 0.2M & 2 & \phantom{0}69\\
       Multi-graded & Gov 200K  & 150 & 0.2M & 2 & 180\\
       Multi-graded & ClueWeb 200K & 150 & 0.2M & 3 &\phantom{0}84 \\
       Binary & MS MARCO 100K  & 97K & 100K & 1 & \phantom{00}1 \\
      Binary &  NQ320K  & 307K & 228K & 1 & \phantom{00}1 \\
        \bottomrule
    \end{tabular}
\end{table}

\subsection{Baselines}
\label{subsec_experiment_baseline}

We first compare our method with traditional retrieval baselines commonly used for document retrieval tasks, including sparse retrieval and dense retrieval methods. The sparse retrieval baselines are:
\begin{enumerate}[label=(\roman*)]
    \item \textbf{BM25} \cite{bm25} is an effective term-based sparse retrieval method, that represents the classical probabilistic retrieval model.
    \item  \textbf{DocT5Query} \cite{doct5query} generates a set of pseudo-queries for each document by a finetuned T5 \cite{raffel2020exploringt5}, and then expand the document with these pseudo-queries.
    \item \textbf{SPLADE} \cite{formal2021splade,formal2021splade-v2} uses a BERT to encode the document into a sparse lexical representation.
\end{enumerate}
The dense retrieval baselines are:
\begin{enumerate}[label=(\roman*)]
    \item \textbf{DPR} \cite{karpukhin2020dense} is a BERT-based dual-encoder model using dense embeddings for text blocks.
    \item \textbf{ANCE} \cite{xiong2020approximate} periodically refreshes the ANN indexer and adpots hard negatives for training a RoBERTa-based dual-encoder model.
    \item \textbf{RepBERT} \cite{zhan2020repbert} is a BERT-based two-tower model. And it takes the in-batch negative sampling technique. RepBERT leverages the representation learning capabilities of BERT to represent the query and document, enhancing dense retrieval performance. 
\end{enumerate}
Further, we explore several advanced GR methods that are trained in a pointwise manner: 
\begin{enumerate}[label=(\roman*)]
    \item \textbf{DSI-Num} \cite{DSI} uses arbitrary unique numbers as docids. And it uses the MLE loss based on query-docid pairs (Eq.~\eqref{eq:retrieval}) and document-docid pairs (Eq.~\eqref{eq:indexing}). 
    \item \textbf{DSI-Sem} \cite{DSI} 
    generates docids by concatenating category numbers obtained through a hierarchical k-means clustering algorithm. This results in similar documents having similar docids. It shares the same training objective as DSI-Num.
    \item \textbf{DSI-QG} \cite{zhuang2022bridgingdsiqg} 
    utilizes pairs of pseudo-queries and docids for indexing. The pseudo-queries are generated conditioned on the document using docT5query~\cite{doct5query}. Similar to DSI-Num, arbitrary unique numbers are used as docids. DSI-QG can be viewed as DSI-Num with data augmentation techniques.
    \item \textbf{NCI} \cite{NCI} 
    replaces the arbitrary unique numbers with semantic structured numbers, similar to DSI-Sem. It uses pairs of pseudo-queries and docids, as well as pairs of leading contents of original documents and docids, to train the model. NCI further designs a prefix-aware decoder, which can distinguish the different meanings of the same number in different positions. NCI can be viewed as the DSI-Sem with data augmentation techniques.
    \item \textbf{GENRE} \cite{genre} retrieves a Wikipedia article by generating its title, specifically designed for the NQ dataset. Due to the absence of titles or incomplete titles in other datasets, we did not experiment with GENRE on those datasets.
    \item \textbf{SEAL} \cite{seal} uses arbitrary n-grams in documents as docids and retrieves documents based on an FM-index during inference.
\end{enumerate}
The GR baselines all optimize indexing (Eq.~\eqref{eq:indexing}) and retrieval (Eq.~\eqref{eq:retrieval}) tasks with MLE, so they can all be considered pointwise approaches.

\subsection{Model variants}
We employ some degraded ListGR models to investigate the effect of our proposed mechanisms:
\begin{enumerate}[label=(\roman*)]
    \item ListGR$_\mathit{pListMLE}$ only trains the model using Eq.~\eqref{eq:training}, and omits the re-training stage. 
    \item ListGR$_\mathit{ListMLE}$ replaces the position-wise loss in ListGR$_\mathit{pListMLE}$ with the ListMLE loss (Eq.~\eqref{eq:listmle}), without considering the position information of docids.
    \item ListGR$_\mathit{Retrain}$ first trains the model using indexing and retrieval loss (Eq.~\eqref{eq:indexing-retrieval}) during the training stage. Then, we perform relevance calibration (Eq.~\eqref{eq:retrain}) over the decoded candidate docid lists during the re-training stage. 
    \item ListGR$_{pListMLE}^{tok}$ first trains the model using Eq.~\eqref{eq:training}, and then re-trains the model with the token-level relevance calibration (Eq.~\eqref{eq:token-level-cases}).
    \item ListGR$_{pListMLE}^{seq}$ first trains the model using Eq.~\eqref{eq:training}, and then re-trains the model with the sequence-level relevance calibration (Eq.~\eqref{eq:retrain-seq}).
    \item ListGR$_{-aug}$ first trains the model (Eq.~\eqref{eq:training}) without augmented data, and then perform relevance calibration (Eq.~\eqref{eq:retrain})  during the re-training stage. 
\end{enumerate}
% \vspace{-2mm}

\subsection{Evaluation metrics}
For datasets with multi-graded relevance labels, i.e., ClueWeb 200K, Gov 200K, and Robust 200K, we perform 5-fold cross-validation to prevent overfitting while maintaining an adequate number of training instances. 
The topic titles are used as queries, and the queries are randomly divided into 5 folds. 
The model parameters are tuned on 4 out of 5 folds, and the remaining fold is used for evaluation. This process is repeated 5 times, with each fold serving as the evaluation set once. 
The final performance is computed by averaging the results from all tested folds. 
The evaluation metrics used in this study are normalized discounted cumulative gain (nDCG$@K$) with $K=\{5,20\}$, expected reciprocal rank (ERR$@20$), and precision at rank 20 (P$@20$), following \cite{guo2016deep,prop,chapelle2011intentERR}.

For datasets with binary relevance labels, i.e., MS MARCO 100K and NQ320K, we adopt the evaluation metrics used in the original DSI model \cite{DSI} and subsequent studies \cite{NCI,zhuang2022bridgingdsiqg,seal}. Specifically, we use mean reciprocal rank (MRR$@K$) with $K=\{3,20\}$ and hit ratio (Hits$@K$) with $K=\{1,10\}$. The performance results are reported on the validation set since the  MS MARCO and NQ leaderboards impose restrictions on submission frequency, following \cite{DSI,prop}.

\subsection{Implementation details}\label{subsec:implementation-details}
\heading{Model architecture} 
Following existing GR works \cite{DSI,NCI,zhuang2022bridgingdsiqg}, we utilize the T5-base model\footnote{\url{https://huggingface.co/t5-base}} as the backbone for ListGR and the baseline models, for a fair comparison. This particular T5-base model is equipped with
a hidden size of 768, a feed-forward layer size of 12, a total of 12 self-attention heads, and a configuration consisting of 12 transformer layers. 

\heading{Baseline implementation}
For BM25, we use the Pyserini \cite{Lin_etal_SIGIR2021_Pyserini} implementation for this baseline. For DSI-Num and DSI-Sem, we re-implement these baselines since the source code is unavailable. For other baselines, we use the publicly available source code for experiments.

\heading{Docid generation}
For the docids used in our work, we leverage semantic structured numbers \cite{DSI,NCI}. 
Specifically, we apply the hierarchical $k$-means algorithm introduced in \cite{DSI} over the document embeddings, which are generated through a 12-layer BERT model with pre-trained parameters, following \cite{DSI,NCI}. 
First, we cluster all documents into 10 clusters. Then, we recursively apply the clustering algorithm for each cluster that consists of more than 100 documents. The result obtained at each level is used as input for the next level, ensuring a well-organized and manageable clustering process. Finally, for each document, all category numbers obtained at each level are concatenated sequentially as its final docid.

\heading{Construction of docid lists}
In the five datasets there exist multiple docids at the same relevance grade with respect to a query. 
During training, we can construct multiple ground-truth docid lists for the query using permutations. 
The length of the list is determined by the highest annotated relevance grade with respect to the query.
Docids within the list are arranged in descending order of relevance grade.

\heading{Hyperparameters}
% ListGR and the reproduced baselines are implemented with PyTorch 1.9.0 and HuggingFace transformers 4.16.2; we re-implement DSI-Num and DSI-Sem, and utilize open-sourced code for other baselines. 
% We use the Adam optimizer with a linear warm-up over the first 10\% steps.
% The learning rate is set to 5e-5, label smoothing to 0.1, weight decay to 0.01, sequence length of documents to 512, maximum training steps to 50K, and batch size to 60.
% We train ListGR on eight NVIDIA Tesla A100 40GB GPUs.
Both ListGR and the reproduced baselines are implemented using HuggingFace transformers 4.16.2. 
For multi-graded relevance datasets, during the training process, we employ the Adam optimizer with a linear warm-up strategy that spans the initial 10\% of steps. Our chosen learning rate is set to 6e-5, with a label smoothing factor of 0.01 and a weight decay rate of 0.01. Furthermore, the sequence length of documents is fixed at 512.
For binary relevance datasets, the hyperparameter settings are as follows: learning rate is 0.001, batch size is 80, and training steps of 100K. We also adopt Adam optimizer with a linear warm-up strategy that spans the initial 200K steps, label smoothing factor of 0.001, and weight decay rate of 0.02.
For all datasets, the maximum number of training steps is capped at 100K, and a batch size of 80 is utilized.
To facilitate the training of ListGR, we make use of eight NVIDIA Tesla A100 40GB GPUs, ensuring efficient computation and faster convergence.

\heading{Training, re-training and inference}
During the training stage, for multi-graded relevance datasets, we set relevance margin $\lambda$ and docid length penalty $\alpha$ (Eq.~\eqref{eq:retrain-seq}) as 0.001 and 0.6, respectively. 
And during the re-training stage, we set $\gamma$ used in Eq.~\eqref{eq:retrain} to 100, and $\beta$ used in Eq.~\eqref{eq:token-level-cases} to 0.002.
For all datasets, to address the limited availability of supervised data, we employ a data augmentation technique that is widely used in existing GR work \cite{{NCI,sun-2023-learning-arxiv,pradeep2023does-scale-genir,chen2023understanding, tang2023semanticenhancedsedsi}}.
% Specifically, we focus on selecting relevant information from each document by choosing the top 3 paragraphs, the top 3 sentences, and randomly sampling 3 entities. These selected paragraphs, sentences, and entities are then associated with their respective docid, effectively building new query-docid pairs. 
% \sout{Furthermore, following \cite{NCI,sun-2023-learning-arxiv}, we generate 5 pseudo-titles for documents with an off-the-shelf title generation model \cite{t5-title-gen} to construct additional query-docid pairs.}
Furthermore, following \cite{NCI,sun-2023-learning-arxiv,pradeep2023does-scale-genir,chen2023understanding}, we generate a set of pseudo-queries for all documents to construct additional query-docid pairs for augmentation. 
Specifically, for MS MARCO 100K, we directly use a publicly trained DocT5query model\footnote{\url{https://github.com/castorini/docTTTTTquery}} on the MS MARCO corpus to generate 20 pseudo-queries for each document. For other datasets, we fine-tune a DocT5query model with labled query-document pairs for them to generate 20 pseudo-queries for training, based on the code\footnote{\url{https://github.com/ArvinZhuang/DSI-QG}} provided in \cite{zhuang2022bridgingdsiqg}.  DSI-QG, NCI, and our ListGR use same pseudo-queries to enhance the training for a fair comparison.
During inference, we construct a decimal trie to constrain the model to decode integers with only 20 beams. 
% construct a prefix trie \cite{genre} for all docids, and adopt constrained beam search to decode docids with 20 beams. s
%\footnote{\url{https://huggingface.co/Michau/t5-base-en-generate-headline}}
\section{Experimental Results}\label{sec:experimental-results}

In this section, we report and analyze the experimental results to demonstrate the effectiveness of the proposed ListGR. We target the following research questions:
\begin{enumerate}[label=(RQ\arabic*),leftmargin=*]
    \item How does ListGR perform compared with strong retrieval baselines across different relevance scenarios?
    \item How do the training and re-training stages of ListGR affect the retrieval performance?
    \item How does ListGR perform in low-resource settings? 
    \item How does the number of relevance grades affect the retrieval performance during training? 
    \item How do the model size and beam size affect the efficiency of retrieval?
    \item Can we better understand how different models perform via some case studies?
\end{enumerate}

\subsection{Baseline comparison}
To answer \textbf{RQ1}, we compare ListGR with several representative traditional retrieval methods and some advanced GR methods, in both multi-graded and binary relevance scenarios.

\subsubsection{\textbf{Results on multi-graded relevance}}
Table \ref{tab:multi-graded-res} shows the performance of ListGR and baselines on multi-graded relevance datasets. We analyze the results in three parts.

\heading{The performance of traditional retrieval baselines} 
\begin{enumerate*}[label=(\roman*)]
\item On the three multi-graded datasets, the dense retrieval baseline ANCE outperforms DPR, RepBERT, and sparse retrieval baselines. The reason may be attributed to its ability to learn rich semantic information, and the strategy of using negative samples  that aids in acquiring stronger discriminative capabilities than sparse retrieval baselines.

\item RepBERT exhibits slightly lower performance than BM25 on Gov 200K and Robust 200K, which aligns with findings reported in previous studies \cite{pipeline4, 10.1145/3477495.3531772costa, lu-etal-2021-lessismore}. The sub-optimal performance of RepBERT in learning effective query and document representations might be primarily attributed to the limited size of the training set available in Gov 200K and Robust 200K. 
\end{enumerate*}

\heading{The performance of generative retrieval baselines}
\begin{enumerate*}[label=(\roman*)]

\item DSI-Sem surpasses the performance of DSI-Num, while SEAL exhibits even higher performance than DSI-Sem. DSI-Num, DSI-Sem and SEAL use random integers, semantic structured clustering numbers, and n-grams from the documents, respectively. 
The integration of docids with stronger semantic associations to the document content can significantly enhance the indexing and retrieval effectiveness of GR. This observation aligns with findings reported in previous studies such as \cite{DSI,seal,genre}.

\item DSI-QG demonstrates superior performance compared to DSI-Num, DSI-Sem, and SEAL, indicating the advantages gained by employing data augmentation techniques that generate additional query-docid pairs. 

\item NCI outperforms DSI-QG due to its use of semantic structured numbers and the presence of the prefix-aware decoder, which effectively distinguishes the meanings of the same numbers in distinct positions within the clustering numerals.

\item NCI and DSI-QG perform slightly better than ANCE, indicating that using pseudo-queries to enhance learning is crucial for GR models. This has been validated in \cite{pradeep2023does-scale-genir} as well.

\end{enumerate*}

\begin{table*}[!h]
    \caption{Experimental results on datasets with multi-graded relevance. $\ast$, $\dagger$, and $\ddagger$ indicate statistically significant improvements over the best performing sparse retrieval baseline SPLADE, dense retrieval baseline ANCE, and generative retrieval baseline NCI, respectively  ($p \leq 0.05$).}
    \label{tab:multi-graded-res}
    \centering
    \setlength{\tabcolsep}{8pt}% column separation
    \renewcommand{\arraystretch}{0.95}%row space 
    \begin{tabular}{l l ccccc}
        \toprule
        & & \multicolumn{2}{c}{\textbf{nDCG}} 
        & \textbf{\phantom{X}P\phantom{X}} 
        & \textbf{ERR}
        \\
        \cmidrule(r){3-4}
        \cmidrule(r){5-5}
        \cmidrule(r){6-6}
        & \multirow{1}{*}{\textbf{Method}}
        & \textbf{$@5$} 
        & \textbf{$@20$}  
        & \textbf{\phantom{X}$@20$\phantom{X}} 
        & \textbf{$@20$}  
        \\
        \midrule
%1
        \multirow{12}{*}{\rotatebox{90}{\textbf{ClueWeb 200K}}}
        &
        BM25 
        & 0.2397 & 0.2568 & 0.3221 & 0.2278 
        \\
        & DocT5query 
        & 0.2542 & 0.2658 & 0.3363 & 0.2328 
        \\
        & SPLADE
        & 0.2588 & 0.2697 & 0.3371 & 0.2357 
        \\
        \cmidrule{2-6}
        & DPR 
        & 0.2672 & 0.2986 & 0.3568 & 0.2806 
        \\
        & ANCE 
        & 0.2694 & 0.3012 & 0.3587 & 0.2815 
        \\
        & RepBERT  
        & 0.2646 & 0.2963 & 0.3520 & 0.2799 
        \\
        \cmidrule{2-6}
        & DSI-Num  
        & 0.1520 & 0.1857 & 0.2182 & 0.1167 
        \\
        & DSI-Sem 
        & 0.1905 & 0.2198 & 0.2563 & 0.1747 
        \\
        & SEAL 
        & 0.2241 & 0.2355 & 0.2725 & 0.1831 
        \\
        & DSI-QG 
        & 0.2765 & 0.2862 & 0.3604 & 0.2825 
        \\
        & NCI 
        & 0.2885 & 0.3058 & 0.3625 & 0.2863 
        \\
        \cmidrule{2-6}
        & ListGR 
        & \textbf{0.3341}\rlap{$^{\ast\dagger\ddagger}$} 
        & \textbf{0.3442}\rlap{$^{\ast\dagger\ddagger}$} 
        & \textbf{0.3704}\rlap{$^{\ast\dagger\ddagger}$} 
        & \textbf{0.2928}\rlap{$^{\ast\dagger\ddagger}$} 
        \\
        \midrule
%2
        \multirow{12}{*}{\rotatebox{90}{\textbf{Gov 200K}}}
        &
        BM25 
        & 0.3712 & 0.3787 & 0.3379 & 0.2398 
        \\
        & DocT5query 
        & 0.3824 & 0.3913 & 0.3435 & 0.2419 
        \\
        & SPLADE
        & 0.3873 & 0.3959 & 0.3486 & 0.2476 
        \\
        \cmidrule{2-6}
        & DPR 
        & 0.3864 & 0.3986 & 0.3584 & 0.2496 
        \\
        & ANCE 
        & 0.3921 & 0.4092 & 0.3605 & 0.2501 
        \\
        & RepBERT  
        & 0.3328 & 0.3443 & 0.3076 & 0.2288 
        \\
        \cmidrule{2-6}
        & DSI-Num  
        & 0.1525 & 0.1588 & 0.1477 & 0.1360 
        \\
        & DSI-Sem 
        & 0.1780 & 0.1469 & 0.1516 & 0.1444  
        \\
        & SEAL 
        & 0.2283 & 0.2053 & 0.1952 & 0.1675 
        \\
        & DSI-QG 
        & 0.3941 & 0.4087 & 0.3635 & 0.2547 
        \\
        & NCI 
        & 0.3986 & 0.4161 & 0.3733 & 0.2629 
        \\
        \cmidrule{2-6}
        & ListGR 
        & \textbf{0.4153}\rlap{$^{\ast\dagger\ddagger}$} 
        & \textbf{0.4368}\rlap{$^{\ast\dagger\ddagger}$} 
        & \textbf{0.3978}\rlap{$^{\ast\dagger\ddagger}$} 
        & \textbf{0.2824}\rlap{$^{\ast\dagger\ddagger}$} 
        \\
        \midrule
%3
        \multirow{12}{*}{\rotatebox{90}{\textbf{Robust 200K}}}
        &
        BM25 
        & 0.3743 & 0.3587 & 0.3456 & 0.2283 
        \\
        & DocT5query 
        & 0.3803 & 0.3617 & 0.3549 & 0.2314
        \\
        & SPLADE
        & 0.3896 & 0.3685 & 0.3573 & 0.2352 
        \\
        \cmidrule{2-6}
        & DPR 
        & 0.3917 & 0.3693 & 0.3588 & 0.2371
        \\
        & ANCE 
        & 0.3952 & 0.3701 & 0.3592 & 0.2393
        \\
        & RepBERT  
        & 0.3608 & 0.3374 & 0.3244 & 0.2097 
        \\
        \cmidrule{2-6}
        & DSI-Num  
        & 0.1649 & 0.1574 & 0.1311 & 0.1205 
        \\
        & DSI-Sem 
        & 0.1887 & 0.1765 & 0.1508 & 0.1566 
        \\
        & SEAL 
        & 0.2209 & 0.2093 & 0.1831 & 0.1769 
        \\
        & DSI-QG 
        & 0.3979 & 0.3723 & 0.3615 & 0.2401 
        \\
        & NCI 
        & 0.4012 & 0.3765 & 0.3678 & 0.2435 
        \\
        \cmidrule{2-6}
        & ListGR  
        & \textbf{0.4284}\rlap{$^{\ast\dagger\ddagger}$} 
        & \textbf{0.3919}\rlap{$^{\ast\dagger\ddagger}$} 
        & \textbf{0.3727}\rlap{$^{\ast\dagger}$} 
        & \textbf{0.2592}\rlap{$^{\ast\dagger\ddagger}$} 
        \\
        \bottomrule
    \end{tabular}
\end{table*}

\begin{table*}[t]
\small
    \caption{Experimental results on datasets with binary relevance. $\ast$, $\dagger$ and $\ddagger$ indicate statistically significant improvements over the best performing sparse retrieval baseline SPLADE, dense retrieval baseline ANCE, and generative retrieval baseline NCI, respectively  ($p \leq 0.05$).}
    \label{tab:binary}
    \centering
    \setlength{\tabcolsep}{8pt}% column separation
    %\renewcommand{\arraystretch}{1.2}%row space 
%    \resizebox{\textwidth}{!}{%
    \begin{tabular}{l ccccc ccccc}
        \toprule
        \multirow{4}{*}{\textbf{Methods}} &  \multicolumn{4}{c}{\textbf{MS MARCO 100K}} &  \multicolumn{4}{c}{\textbf{NQ320K}}\\
        \cmidrule(r){2-5}
        \cmidrule{6-9}

        & \multicolumn{2}{c}{\textbf{MRR}} & 
        \multicolumn{2}{c}{\textbf{Hits}}
        & \multicolumn{2}{c}{\textbf{MRR}} & 
        \multicolumn{2}{c}{\textbf{Hits}}\\
        \cmidrule(r){2-3}
        \cmidrule(r){4-5}
        \cmidrule(r){6-7}
        \cmidrule(r){8-9}

        & \textbf{$@3$} & \textbf{$@20$} & \textbf{\phantom{1}$@1$\phantom{1}} & \textbf{$@10$}  & \textbf{$@3$} & \textbf{$@20$} & \textbf{\phantom{1}$@1$\phantom{1}} & \textbf{$@10$}  \\
        \midrule

         BM25 &  0.3884 & 0.4157 & 0.4912 & 0.5572  & 0.2849 & 0.4426 & 0.2927 & 0.6015  \\
         DocT5query& 0.4053 & 0.4376 & 0.5029 & 0.5741  & 0.3641 & 0.4825 & 0.3913 & 0.6972\\
        SPLADE& 0.4164 & 0.4454 & 0.5095 & 0.5813  & 0.4467 & 0.7036 & 0.4982 & 0.7835\\
        \midrule
        DPR& 0.4212 & 0.4598 & 0.5214 & 0.6124  & 0.4792 & 0.7583 & 0.5024 & 0.8042\\
        ANCE& 0.4235 & 0.4601 & 0.5327 & 0.6267  & 0.4821 & 0.7622 & 0.5183 & 0.8149\\
         RepBERT  &0.4202 & 0.4571 & 0.5183 & 0.6052 & 0.4589 & 0.7154 & 0.4835 & 0.7981\\
        \midrule
        
         DSI-Num & 0.1348 & 0.1353 & 0.1264 & 0.1218 & 0.1815 & 0.3785 & 0.2214 & 0.4184 \\
         DSI-Sem & 0.2278 & 0.2209 & 0.2123 & 0.2714 & 0.2198 & 0.4248 & 0.2793 & 0.5763 \\
        GENRE  & -- &	-- &	 --	& --	& 	0.3543&	0.6218&	0.3942&	0.7061 \\
         SEAL & 0.3299 & 0.3771 & 0.3721 & 0.5397 & 0.3672 & 0.6398 & 0.4173 & 0.7289  \\
         DSI-QG &  0.4276 & 0.4524 & 0.5273 & 0.6285 & 0.5834 & 0.7592 & 0.6349 & 0.8236   \\
         NCI  &  0.4359 & 0.4638 & 0.5362 & 0.6396 & 0.5952 & 0.7641 & 0.6425 & 0.8332  \\
        \midrule

        ListGR &
     \textbf{0.4656}\rlap{$^{\ast\dagger\ddagger}$} & \textbf{0.4901}\rlap{$^{\ast\dagger\ddagger}$} & \textbf{0.5576}\rlap{$^{\ast\dagger\ddagger}$} & \textbf{0.6471}\rlap{$^{\ast\dagger\ddagger}$} & \textbf{0.6019}\rlap{$^{\ast\dagger\ddagger}$} & \textbf{0.7723}\rlap{$^{\ast\dagger\ddagger}$} & \textbf{0.6593}\rlap{$^{\ast\dagger\ddagger}$} & \textbf{0.8412}\rlap{$^{\ast\dagger\ddagger}$} 
 \\

        \bottomrule
    \end{tabular}
%    }
\end{table*}

\heading{The performance of ListGR} By adopting a listwise approach in which lists of docids are used as ``instances'' in learning, ListGR achieves significantly better performance than existing generative retrieval baselines that work in a pointwise manner. Specifically, on the ClueWeb 200K dataset, ListGR outperforms NCI by 15.8\% in terms of nDCG@5. On the Gov 200K dataset, ListGR surpasses NCI by 7.4\% in terms of ERR@20. On the Robust 200K dataset, ListGR surpasses NCI by 6.8\% in terms of nDCG@5. Furthermore, this outcome suggests that the inclusion of additional relevance levels within the annotated data, such as ClueWeb 200K, yields substantial benefits for ListGR. By incorporating more comprehensive relevance information, ListGR can effectively learn and accurately assess the relevance order among the docid list.

\subsubsection{\textbf{Results on binary relevance}}
For the binary relevance datasets, where the positional weight ($\alpha(i) = 2^{n-i}-1$) of relevant docids is zero, the training stage only utilizes the indexing and retrieval loss (Eq.~\eqref{eq:indexing-retrieval}). Based on this, the trained model undergoes relevance calibration. Table~\ref{tab:binary} shows the performance of ListGR and baselines on binary relevance datasets.
We observe the following:
\begin{enumerate*}[label=(\roman*)]
    \item The three dense retrieval baselines outperform sparse retrieval baselines. This could be attributed to the availability of abundant labeled query-document pairs in these two datasets. It helps dense models learn dense representations and captures the semantic relationship between queries and documents.
    \item DSI-Num and DSI-Sem perform worse than dense retrieval baselines, e.g., RepBERT, DPR and ANCE on both binary relevance datasets. This suggests that learning both indexing and retrieval tasks simultaneously through these two types of docids and MLE is still challenging.  
    \item SEAL shows better performance than vanilla DSI methods, i.e., DSI-Num and DSI-Sem. The reason might be that SEAL uses n-grams from the documents as docids. This type of docid contains more explicit semantics, which helps the model learn better than numeric docids.
    \item Moreover, both DSI-QG and NCI outperform SEAL, DSI-Num and DSI-Sem, indicating that data augmentation methods, such as transforming documents into pseudo-queries for learning, contribute significantly to the improvement.
    \item ListGR outperforms the best-performing GR baseline, NCI, on both binary relevance datasets. Specifically, ListGR achieves improvements of 6.8\% in terms of MMR@3 on MS MARCO 100K. This indicates that relevance calibration has the ability to correct inappropriate ordering of docid lists generated by beam search decoding.
\end{enumerate*}

\begin{table*}[h]
    \caption{Ablation analysis of ListGR with its variants on multi-graded relevance datasets. $\ast$ indicates statistically significant improvements over all the corresponding variants ($p \leq 0.05$).}
    \label{tab:ablation}
    \centering
    \setlength{\tabcolsep}{3pt}% column separation
    \renewcommand{\arraystretch}{1.4}%row space 
    \begin{tabular}{l l cccc}
        \toprule
        & & \multicolumn{2}{c}{\textbf{nDCG}} 
        & \textbf{\phantom{X}P\phantom{X}} 
        & \textbf{ERR}  
        \\
        \cmidrule(r){3-4}
        \cmidrule(r){5-5}
        \cmidrule(r){6-6}
        \multirow{1}{*}{\textbf{Method}} &
        & \textbf{$@5$} 
        & \textbf{$@20$}  
        & \textbf{\phantom{X}$@20$\phantom{X}} 
        & \textbf{$@20$}  
        \\
        \midrule
%1
        \multirow{7}{*}{\rotatebox{90}{\textbf{ClueWeb 200K}}}
        &
        ListGR$_\mathit{pListMLE}$ 
        & 0.3087 & 0.3205 & 0.3618 & 0.2887 
        \\
        & ListGR$_\mathit{ListMLE}$ 
        & 0.2947 & 0.3114 & 0.3609 & 0.2874 
        \\
        & ListGR$_\mathit{Retrain}$ 
        & 0.2961 & 0.3156 & 0.3686 & 0.2881 
        \\
        & ListGR$_{pListMLE}^{tok}$ 
        & 0.3224 & 0.3302 & 0.3641 & 0.2894 
        \\
        & ListGR$_{pListMLE}^{seq}$
        & 0.3252 & 0.3331 & 0.3668 & 0.2908 
        \\ 
        & ListGR$_{-aug}$ 
        & 0.2713 & 0.2811 & 0.3509 & 0.2746 
        \\
        \cmidrule{2-6}
        & ListGR 
        & \textbf{0.3341}\rlap{$^{\dagger}$} 
        & \textbf{0.3442}\rlap{$^{\dagger}$} 
        & \textbf{0.3704} 
        & \textbf{0.2928} 
        \\
        \midrule
%2        
        \multirow{7}{*}{\rotatebox{90}{\textbf{Gov 200K}}}
        & 
        ListGR$_\mathit{pListMLE}$ 
        & 0.3998 & 0.4214 & 0.3842 & 0.2765 
        \\
        & ListGR$_\mathit{ListMLE}$ 
        & 0.3991 & 0.4185 & 0.3787 & 0.2685 
        \\
        & ListGR$_\mathit{Retrain}$ 
        & 0.3995 & 0.4192 & 0.3818 & 0.2716 
        \\
        & ListGR$_{pListMLE}^{tok}$ 
        & 0.4062 & 0.4256 & 0.3871 & 0.2782 
        \\
        & ListGR$_{pListMLE}^{seq}$
        & 0.4094 & 0.4288 & 0.3919 & 0.2809 
        \\ 
        & ListGR$_{-aug}$ 
        & 0.3551 & 0.3731 & 0.3036 & 0.2204 
        \\
        \cmidrule{2-6}
        & ListGR 
        & \textbf{0.4153} 
        & \textbf{0.4368}\rlap{$^{\dagger}$} 
        & \textbf{0.3978} 
        & \textbf{0.2824} 
        \\
        \midrule
%3        
        \multirow{7}{*}{\rotatebox{90}{\textbf{Robust 200K}}}
        &
        ListGR$_\mathit{pListMLE}$ 
        & 0.4074 & 0.3798 & 0.3694 & 0.2483
        \\
        & ListGR$_\mathit{ListMLE}$ 
        & 0.4057 & 0.3778 & 0.3685 & 0.2456
        \\
        & ListGR$_\mathit{Retrain}$ 
        & 0.4068 & 0.3783 & 0.3689 & 0.2471
        \\
        & ListGR$_{pListMLE}^{tok}$ 
        & 0.4145 & 0.3826 & 0.3697 & 0.2498
        \\
        & ListGR$_{pListMLE}^{seq}$
        & 0.4193 & 0.3851 & 0.3705 & 0.2559
        \\ 
        & ListGR$_{-aug}$ 
        & 0.3528 & 0.3026 & 0.2971 & 0.2066
        \\
        \cmidrule{2-6}
        & ListGR 
        & \textbf{0.4284}\rlap{$^{\dagger}$} 
        & \textbf{0.3919}\rlap{$^{\dagger}$} 
        & \textbf{0.3727} 
        & \textbf{0.2592} 
        \\
        \bottomrule
    \end{tabular}%
\end{table*}

\subsection{Ablation study}
In this section, to answer \textbf{RQ2}, we conduct an ablation analysis on three multi-graded relevance datasets to quantitatively assess the impact of each component in ListGR; see Table~\ref{tab:ablation}. For the binary relevance datasets, the training stage lacks listwise loss, so that ListGR and ListGR$_\mathit{Retrain}$ are the same in this setting; therefore, we did not analyze the performance on  binary relevance datasets in this context.
We have the following observations: 

\heading{Listwise loss}
\begin{enumerate*}[label=(\roman*)]
    \item ListGR$_\mathit{Retrain}$, only using the re-training stage leads to significantly lower performance than ListGR. Additionally, in the training stage, ListGR$_\mathit{pListMLE}$ and ListGR$_\mathit{ListMLE}$ combining a listwise loss with an indexing and retrieval loss improves the retrieval performance over NCI (in Table~\ref{tab:multi-graded-res}).
     These results indicate that modeling the ranked docid list explicitly is crucial for better retrieval performance, as using MLE alone does not capture the relationships between docids. 
    \item ListGR$_\mathit{pListMLE}$ performs better than ListGR$_\mathit{ListMLE}$, highlighting the importance of position weights in ranking, aligning with the observations in \cite{lan2014positionplistmle}.
    \item ListGR$_{-aug}$ significantly outperforms SEAL (in Table \ref{tab:multi-graded-res}). It demonstrates that our listwise approach, even without data augmentation, can assist the GR model in learning stronger discriminative abilitty for relevance.
\end{enumerate*}

\heading{Relevance calibration} 
\begin{enumerate*}[label=(\roman*)]
    \item By removing relevance calibration, ListGR$_\mathit{pListMLE}$ and ListGR$_\mathit{ListMLE}$ have a significant drop in performance compared to ListGR. This suggests that beam search decoding has an impact on the inference effectiveness of GR.
    \item Additionally, both ListGR$_{pListMLE}^{tok}$ and ListGR$_{pListMLE}^{seq}$, built upon ListGR$_\mathit{pListMLE}$, show improved performance. This indicates that further relevance calibration to candidate docids is essential.
    \item  Furthermore, we observe that the performance of ListGR$_{pListMLE}^{tok}$ and ListGR$_{pListMLE}^{seq}$ is similar, suggesting that both sequence-level and token-level relevance calibration are crucial for the GR model.
\end{enumerate*}
These results demonstrate that adjusting the generation probabilities of docids in the candidate docid list generated by the trained model contributes to generating more accurate ranking positions in the list.

\subsection{Low-resource settings}
In this section, to answer \textbf{RQ3}, during training, 
 we simulate a low-resource retrieval scenario by randomly sampling a fixed and limited number of queries from the training set.
More specifically, for the purpose of comparing ListGR and NCI, we randomly sample 15, 30, 45, and 60 queries from the ClueWeb 200K, Gov 200K, and Robust 200K datasets. For the MS MARCO 100K and NQ320K datasets, we randomly sample 2K, 4K, 6K, and 8K queries.

\begin{figure}[t]
 \centering
 \includegraphics[width=\textwidth]{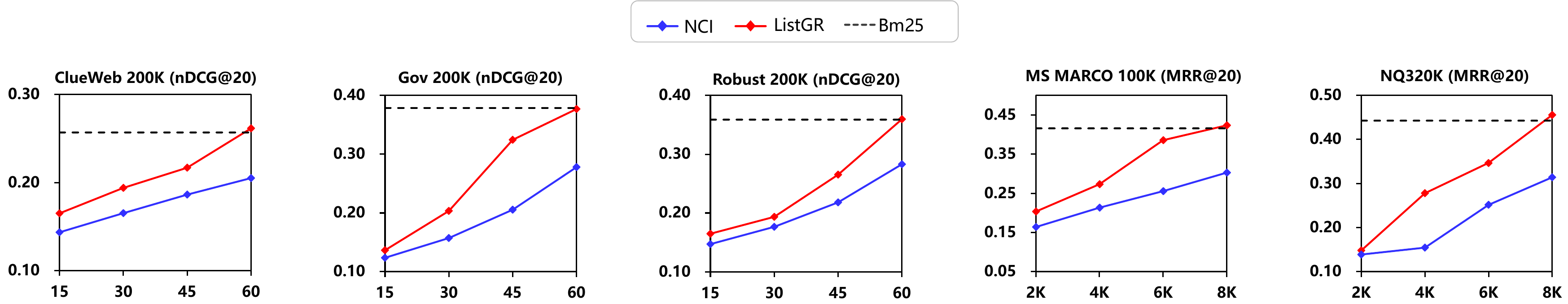}
 \caption{Training with limited supervision data. The x-axis indicates the number of training queries.}
 \label{fig:low-resources}
 \vspace{-2mm}
\end{figure}

Based on Figure~\ref{fig:low-resources}, we observe the following:
\begin{enumerate*}[label=(\roman*)]
    \item On multi-graded relevance datasets, ListGR outperforms NCI, which suggests that ListGR is capable of modeling the relevance of docid lists using limited information.
    \item Similarly, on binary relevance datasets, ListGR achieves better performance than NCI, indicating that the relevance calibration stage can further enhance the model's ability to recognize the relevance order of docids within the list, even under the pointwise training objective.
    \item ListGR exhibits superior performance compared to a strong BM25 baseline on most datasets. For example, on the ClueWeb 200K dataset, ListGR achieves comparable performance with 58 queries in terms of nDCG@20, while on the MS MARCO 100K dataset, ListGR performs well with only 8\% queries, i.e., 7.8K queries in terms of MRR@20.
\end{enumerate*}

\subsection{Analysis of the relevance grades}
To answer \textbf{RQ4}, we conduct an analysis by controlling the number of relevance grades employed in the listwise loss during the training phase. This investigation assesses the influence of different numbers of relevance grades on the performance of ListGR. 

Specifically, we conduct experiments on the ClueWeb 200K dataset using three, two, and one relevance grades in Eq.~\eqref{eq:listwise-GR}, respectively. 
For the case of using two relevance grades, we further divide it into three scenarios: using 2- and 3-grades, using 1- and 3-grades, and using 1- and 2-grades for training. 
Using only one relevance grade data is equivalent to training with MLE alone (Eq.~\eqref{eq:indexing-retrieval}), which has the same effect as ListGR$_\mathit{Retrain}$. During testing, we uniformly use the original  testing set consistently across all the aforementioned scenarios.

\begin{figure}[t]
 \centering
 \includegraphics[width=0.8\textwidth]{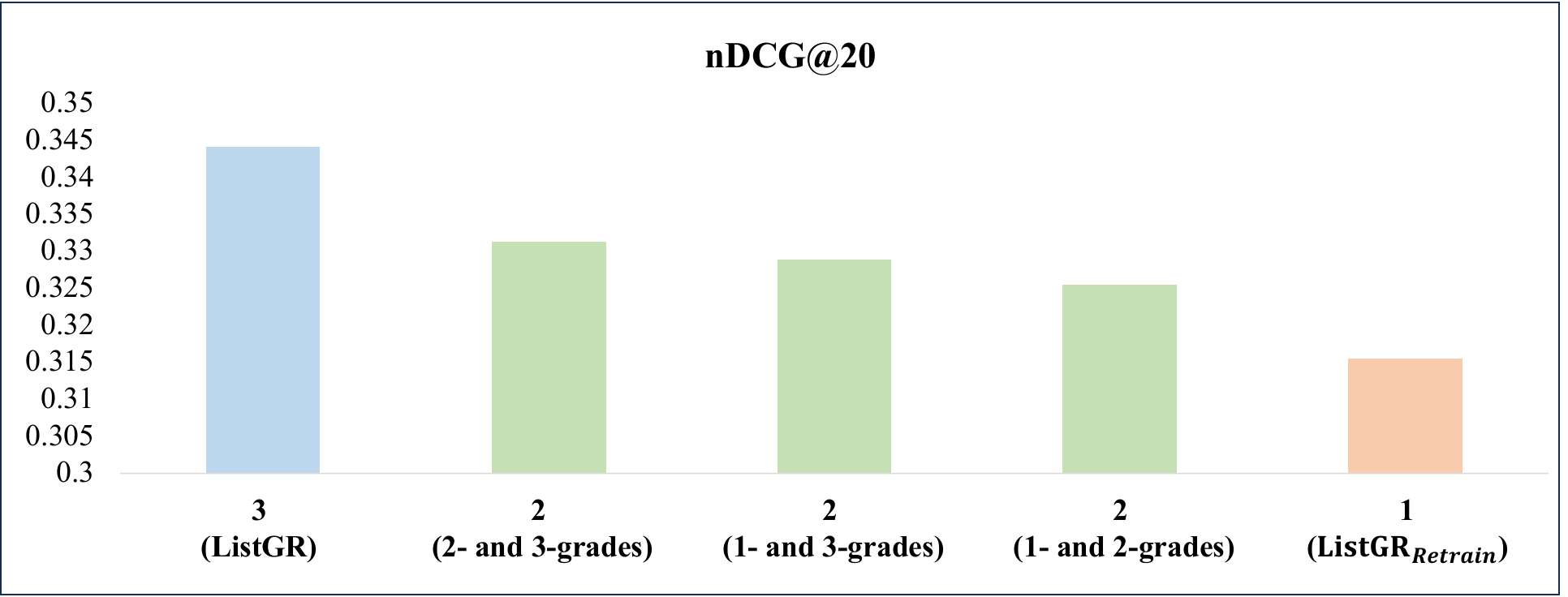}
 \caption{During the training stage, different numbers of relevance grades in the ClueWeb 200K dataset are used in the listwise loss (Eq.~\eqref{eq:listwise-GR}). The x-axis represents the number of relevance grades used, indicated in parentheses as the combination of the relevance grades or the corresponding model names.}
 \vspace{-2mm}
 \label{fig:relevance-grades}
\end{figure}

Based on Figure \ref{fig:relevance-grades}, we observe the following:
\begin{enumerate*}[label=(\roman*)]
    \item On the same dataset, increasing the number of relevance grades used in the listwise loss (Eq.~\eqref{eq:listwise-GR}) during the training stage leads to better performance. For example, using three relevance grades (blue bar) yields a higher nDCG@20 value than using two (green bars) or one (orange bar) relevance grades only. This could be because providing more relevance labels allows the model to learn more comprehensive and fine-grained differences in relevance. 
    \item Among the scenarios using two relevance levels, incorporating 3-graded data results in better performance. For instance, both scenarios using 2- and 3-grades, and using 1- and 3-grades have higher nDCG@20 values than the scenario using 1- and 2-grades. This suggests that docids with higher relevance grades may carry more importance in the list, and learning these docids contributes to better docid list generation.
\end{enumerate*}

\subsection{Efficiency analysis}
To answer \textbf{RQ5}, we analyze the efficiency using an NVIDIA A100-40GB GPU. It is important to note that the inference speed of ListGR is influenced by two factors: model capacity and beam size. In order to provide comprehensive insights, following \cite{NCI}, we have included the latency and throughput measures for various settings in Table~\ref{tab:efficiency}. 
Specifically, latency refers to the time it takes for a retrieval model to process a query.
And throughput represents the speed at which a retrieval model can process a certain number of queries within a second.
Based on the ClueWeb 200K dataset, for latency, we randomly sampled multiple batches of queries, measured the total time for inference, and then divided it by the number of queries to obtain latency. For throughput, we also randomly sampled multiple batches of queries, measured the average number of queries inferenced in 1 second, and obtained the throughput.

\begin{table}[h]
    \caption{Efficiency analysis. According to two important factors, namely model size and beam size, ListGR demonstrates encouraging performance in terms of latency and throughput.}
    \label{tab:efficiency}
    \centering
    \setlength\tabcolsep{10pt}
    \begin{tabular}{lccc}
         \toprule
        \textbf{Model size} & \textbf{Beam size}  & \textbf{Latency (ms)} & \textbf{Throughput (queries/s)} \\
        \midrule
        Small & \phantom{0}10 & \phantom{0}76.38	&59.73\\
        Base & \phantom{0}10 & 112.56	&54.28\\
        Large & \phantom{0}10 & 180.64	&45.53\\
        \midrule 
        Small & 100& 218.25&	\phantom{0}7.81\\
        Base & 100& 264.07	& \phantom{0}5.32\\
        Large & 100& 357.81&	\phantom{0}4.16\\
        \bottomrule
    \end{tabular}    
\end{table}

\noindent In terms of latency and throughput, ListGR demonstrates promising performance for certain near-real-time applications. The latency of ListGR is comparable to that of DSI \cite{DSI} when using the same model size and beam size, as both approaches employ beam search with transformer decoders. Similar phenomena is observed in \cite{NCI}.
BM25 has higher retrieval efficiency, but due to a lack of semantic matching, its retrieval performance is lower. RepBERT has lower efficiency because it performs brute-force search based on dense vectors, making it more time-consuming.

\subsection{Case study}
To answer \textbf{RQ6}, we perform case studies from two perspectives. First, we scrutinize the docid lists generated by various methods for a given query. Second, we employ visualization techniques to assess the representations of the query and its candidate documents.

\heading{Textual analysis}
We take a sample from the test set of ClueWeb 200K and compare the top-5 docid lists predicted by ListGR and NCI. Since both models use semantic structured numbers as docids, we also summarize the topics of the corresponding documents for better understanding and analysis of the differences; see Table~\ref{tab:case-study-with-nci}.
Given the query ``horse hooves'' (QID: 51), docids predicted by ListGR align with their respective relevance labels. However, NCI fails to predict any docids with a relevance level of 3 and struggles to distinguish the relative order of docids with relevance levels 2 and 1. This indicates that the objective of modeling the docid list in ListGR contributes to generating accurate and high-quality docid lists in GR.

\begin{table}[t]
\small
    \caption{An example from the ClueWeb 200K dataset, given the query ``horse hooves,'' which has relevant docids with three different grades, ListGR and NCI return the top-5 beams. We also present the corresponding topics and relevance labels of these predicted docids.}
    \label{tab:case-study-with-nci}
    \centering
    \renewcommand{\arraystretch}{1}%row space 
    \setlength\tabcolsep{4pt}
    \begin{tabular}{c rlc rlc}
         \toprule
        &  \multicolumn{3}{c}{\textbf{ListGR}} &  \multicolumn{3}{c}{\textbf{NCI}} \\
        \cmidrule(r){2-4}
        \cmidrule(r){5-7}

        \textbf{\#Rank} & Docid & Topic  & Label & Docid & Topic & Label\\
        \midrule
        1& 95573 & Taking Care Of Horse's Hooves& 3& 716310 & Horse Care Products &1\\
        2& 582003 & The Barefoot Horse &2 &777805 & Horse Information &1\\
        3&729007 & Steel Horseshoes &2& 729007 & Steel Horseshoes &2\\
        4& 729707 & All About Horses &2& 729707 & All About Horses &2\\
        5& 716310 & Horse Care Products &1& 777711& Pap test & 0\\
        \bottomrule
    \end{tabular}   
    \vspace{-3mm}
\end{table}

\begin{figure}[h]
 \centering
 \includegraphics[width=0.8\textwidth]{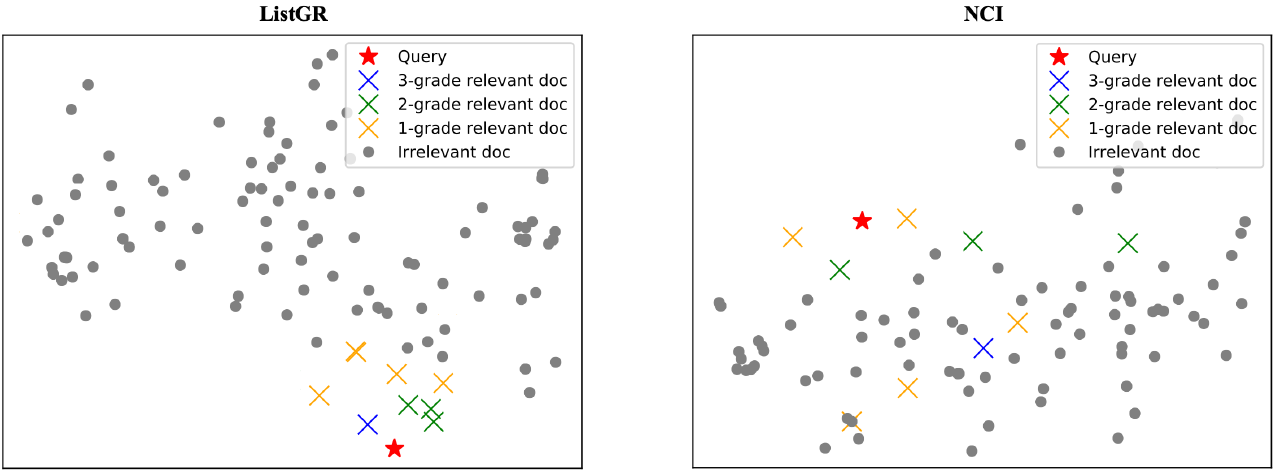}
 \caption{t-SNE plots of query and document representations for ListGR and NCI. The representations are the output of the encoder of ListGR and NCI.}
 \label{fig:visual}
\end{figure}

\heading{Visual analysis}
To deepen our understanding of ListGR, we 
employ t-SNE \cite{van2008visualizingtsne} for visualizing the distributions of query and document representations in the semantic space. Expanding on the previous query sample, we create a t-SNE plot to compare the representations of the sampled query and its top-100 candidate documents generated by the encoder output of ListGR and the best-performing GR baseline, NCI \cite{NCI}.

As shown in Figure \ref{fig:visual}, for ListGR, documents with higher relevance levels are closer to the query, while irrelevant documents are located far away. In the case of NCI, 
1-grade relevant documents are closest to the query, while 2- and 3-grade  relevant  documents are much further away. This demonstrates that ListGR has the ability to differentiate the relevance of docids in a more fine-grained manner in the docid list.

\section{Related Work}
\label{sec:related-work}

In this section, we review related work, including the traditional document retrieval, pre-trained language models, and generative retrieval.

\subsection{Traditional document retrieval}
Document retrieval has traditionally followed an ``index-retrieve'' paradigm, where documents are indexed and then retrieved based on a query. This paradigm has resulted in two main approaches to document retrieval, namely sparse retrieval and dense retrieval.

\subsubsection{\textbf{Sparse retrieval.}}
Sparse retrieval methods represent queries and documents using sparse vectors. These methods rely on exact matching to compute similarity scores between queries and documents. In sparse retrieval, the focus is on identifying the presence or absence of specific query terms within documents.
Two typical methods in this category are BM25 \cite{bm25} and the query likelihood model \cite{lavrenko2017relevancequerylike}.
BM25 takes into account factors such as document length, term frequency, and inverse document frequency to rank documents based on the occurrence of query terms within each document. The query likelihood model \cite{lavrenko2017relevancequerylike}, on the other hand, 
leverages a generative model and estimates the probability of generating the query terms given a document. Documents are then ranked based on their likelihood of generating the query.
However, these approaches solely consider statistical information and do not incorporate semantic information. To overcome this limitation, several studies \cite{pipeline2, bendersky2010learning, deepCT, bendersky2012effective, deepTR, bendersky2011parameterized} have utilized word embeddings to reweight the importance of terms. For example, HDCT \cite{dai2020contextHDCT}
focuses on long documents.
It first utilizes BERT to generate contextual term representations, which are then used to estimate passage-level term weights. Subsequently, these passage-level term weights are aggregated using a weighted sum to obtain document-level term weights.
DeepTR \cite{deepTR} constructs a feature vector for query terms and employs a regression model to map these feature vectors to the ground truth weights of terms. 
% Additionally, DeepCT utilizes contextualized representations learned by BERT \cite{bert} as term weights.

\heading{Limitations}
Sparse retrieval methods offer computational efficiency due to their reliance on exact matching. They are particularly useful in large-scale retrieval scenarios where the number of documents is substantial. However, these methods often lack the ability to capture semantic relationships and contextual information between query terms and documents, which can limit their retrieval performance.

\subsubsection{\textbf{Dense retrieval.}}
Unlike sparse retrieval methods that rely on exact matching, which gives rise to the vocabulary mismatch problem \cite{zhao2010term,furnas1987vocabulary} dense retrieval focuses on capturing semantic relationships and contextual information \cite{zhan2020repbert,luan2021sparse,hofstatter2021efficiently,xiao2022retromae,zeng2022curriculum}. It represents both queries and documents as continuous, dense vectors in a high-dimensional semantic space, to calculate similarity, i.e., using the dot product or cosine similarity as the relevance score. 

To enhance the efficiency of dense retrieval, approximate nearest neighbor search methods \cite{bentley1975multidimensional, beis1997shape} are employed. These methods accelerate the retrieval process by finding approximate nearest neighbors instead of exact matches.
In addition, numerous pre-trained models and techniques have been leveraged to further improve the performance of dense retrieval \cite{nie2020dc, khattab2020colbert, chang2020pre, guu2020retrieval, abnarexploring, lee2020contextualized}. For instance,
DC-BERT \cite{nie2020dc} employs dual BERT encoders.
In the lower layers, an online BERT encoder is responsible for encoding the query once, while an offline BERT encoder pre-encodes all the documents and stores their term representations in a cache. The obtained contextual term representations are fed into high-layer transformer interaction, initialized by the last few layers of the pre-trained BERT.
These approaches take advantage of pre-trained models and advanced techniques to enhance the quality of dense retrieval and 
can capture more nuanced and subtle semantic relationships between words and phrases in queries and documents,  which are often challenging for sparse retrieval methods.
To enhance performance, the ranking module is also often leverages \cite{zeng2022curriculum} . In this work, we focus only on the ``index-retrieve'' stage, leaving ranking enhancement for future work. To improve efficiency, approximate nearest neighbor algorithms \cite{ge2013optimized,jegou2010product,xiong2020approximate} and various sampling methods \cite{hofstatter2021efficiently,yang2020mixed} have been proposed.

\heading{Limitations}
Despite the promising performance of the ``index-retrieve'' paradigm in dense retrieval, there are limitations that need to be addressed:
\begin{enumerate*}[label=(\roman*)]
    % \item  The vanilla dense retrieval method needs additional fine-grained rerankers \cite{prop, ma2021bprop, chang2020pre} for achieving optimal performance in real-world applications. The current approach of using multiple heterogeneous modules, each optimized for a different objective, often falls short of delivering the desired results.
    \item During training, a query encoder and a document encoder are utilized to generate representations for the query and the document, respectively. However, the independence of these encoders restricts the depth of interactions between the representations, thus posing a risk of missing information. Furthermore, the discrete modules in the system cannot be optimized in an end-to-end manner, resulting in sub-optimal performance.
    \item During inference, the query is required to search for relevant documents across the entire corpus. Although efficiency-enhancing strategies are available, such as approximate nearest neighbor search, these methods may sacrifice some semantic information in the process.
\end{enumerate*}
These limitations highlight the need for further advancements to explore more efficient methods that can retain important semantic information during the retrieval process.

\subsection{Pre-trained language models}
Pre-trained models have revolutionized natural language processing tasks by leveraging large-scale unsupervised training on vast amounts of text data, with pre-training and fine-tuning techniques \cite{abnarexploring,weifinetuned,lilarge,korbakreinforcement,NEURIPS2021_0cd6a652,NEURIPS2021_22b1f2e0,NEURIPS2021_86b3e165,balunovic2022lamp,korbak2022controlling,lang2022co}.
Usually, these models are trained to learn contextualized representations of words, sentences, or documents, which capture rich semantic and syntactic information.
Pre-trained models can be broadly classified into two categories, namely discriminative models and generative models.

\subsubsection{\textbf{Discriminative pre-trained models.}}
Discriminative pre-trained models are primarily designed for tasks that involve classification, regression, or any other form of prediction.
Examples of discriminative models include BERT \cite{bert}, RoBERTa \cite{liu2019roberta}, and SpanBERT \cite{joshi2020spanbert}.
Further, they are widely used in IR, for example, BERT is used to re-weight term weights \cite{deepCT,dai2020contextHDCT,deepTR} in spare retrieval. Furthermore, dual BERT architectures are used to learn dense query and document representations to support fine-grained semantic interaction \cite{khattab2020colbert,nie2020dc, guu2020retrieval, abnarexploring, lee2020contextualized}
in dense retrieval.
To bridge the gap between general pre-trained language models and downstream retrieval tasks, some studies \cite{prop,ma2021bprop,xiao2022retromae,ma2021pre,lee2019latent} have proposed specialized pre-training tasks for the retrieval.

\subsubsection{\textbf{Generative pre-trained models.}}
In addition to discriminative pre-trained models, there has been a growing focus on generative pre-trained models and techniques \cite{lamprier2022generative,bae2022noisy,zhang2021understanding,hudson2021generative,lin2021generative,zhang2020pegasus,raffel2020exploringt5} for text generation. 
Generative models typically use autoregressive modeling techniques, such as language modeling, where they predict the next word or token in a sequence based on the previous context. Examples of generative models include GPT \cite{2022Traininggpt}, BART \cite{Lewis2019BARTDS}, T5 \cite{raffel2020exploringt5}. 
They have also been researched and applied in IR, for example, T5 is utilized to generate queries for a document. These synthetic queries are then appended to the original documents, creating an ``expanded document'' to enhance document retrieval \cite{doct5query}.
And in \cite{nogueira2020documentgptranking},
given a document, the conditional likelihood of generating queries using GPT serves as the relevance score, which is used for ranking.
And \citet{dos2020beyond} propose that, given a query and document, T5 concatenates them as input and produces either a ``True'' or ``False'' token as output; if the query is relevant to the document, it outputs ``True'' and proceeds to calculate the generation probability as the relevance score; if the query is irrelevant, it outputs ``False''.

\heading{Limitations}
While these explorations with generative models have shown some improvements in information retrieval, some work still revolve around matching queries with documents. 
This method faces limitations when it comes to dealing with a substantial volume of documents, and it incurs a high computational burden. 

% This method has limitations in terms of handling a large number of documents and imposes a high computational cost. 

\subsection{Generative retrieval}
In order to further develop the capabilities of generative models, a new retrieval paradigm based on generative models has been proposed, called generative retrieval (GR) \cite{modelBased}. 
GR aims to directly generate relevant docids for a qiven query.
GR methods parameterize the corpus information, by replacing the traditional external index by a training process that learns the mapping from documents to their corresponding document identifiers (docids). 
% Instead of relying on an index, a single model is trained to directly generate a list of relevant docids using its learned parameters.
Building upon this framework, researchers have proposed various approaches \cite{genre,seal,chen2022corpusbrain,lee2022contextualized,DSI,NCI,chen-2023-unified,sun-2023-learning-arxiv,li2023multiview,zeng2023scalable,ren2023tome}.
GR needs to learn a Seq2Seq model that address two key tasks simultaneously, namely indexing and retrieval.

\subsubsection{\textbf{Indexing task.}}
In \ac{GR} this task  is aimed at establishing associations between documents and docids.
% In current GR methods, two primary approaches are employed to represent docids:
% \begin{enumerate*}[label=(\roman*)]
% \item Using arbitrary unique integers without explicit semantic connections to the corresponding documents \cite{DSI}, or
% \item Utilizing strings that carry semantic associations with the documents, often obtained through techniques like hierarchical k-means clustering \citep{DSI, NCI}.
% \end{enumerate*}
% Studies have shown that incorporating semantic associations between docids and documents improves the retrieval process \citep{DSI,NCI, genre, seal}. 
For the document identifiers, in addition to the two primary approaches described in Section~\ref{sec:preliminaries} -- arbitrary unique integers and structured semantic numbers --, there are other types of identifiers.
Document titles have garnered considerable attention as they possess inherent semantic relevance \citep{chen2022corpusbrain}. 
However, methods that use document titles heavily rely on the availability of specific document metadata, limiting their applicability. To address this limitation, some approaches have explored using all n-grams within a passage as its docid \citep{seal}. 
Moreover, the utilization of pseudo-queries generated from the documents as docids has shown significant improvements in retrieval performance \cite{tang2023semanticenhancedsedsi}. This is because such docids can represent key information about the documents to some extent.
\citet{ren2023tome} leverage tokenized URLs as docids, which may contain key phrases of documents.
To provide a more comprehensive representation of the document's information, \citet{li2023multiview} use multiple docids to represent a single document.

To encode the entire corpus, existing approaches primarily employ a Seq2Seq framework, where the original document is taken as input, and the corresponding docid is generated as the output. In this way, the index is embedded within the model parameters, and indexing becomes an integral part of the model training process. Building on~\citep{DSI}, we adopt a straightforward input-to-target approach, explicitly associating document tokens with their corresponding docids.

\subsubsection{\textbf{Retrieval task.}}
In \ac{GR} this task focuses on mapping queries to relevant docids.
Current GR models typically employ a teacher forcing approach \citep{1998Ateacher,hao2022teacher,lin2022r}, maximizing the likelihood of the output sequence conditioned on the input query. If a query has multiple relevant docids, it learns multiple query-docid pairs.

Building upon this blueprint, the first exploration of the GR paradigm was undertaken by GENRE \cite{genre}. GENRE utilized the unique titles of Wikipedia articles as document identifiers and employed the BART model \cite{Lewis2019BARTDS} to directly generate a list of relevant article titles for a given query using constrained beam search, with a prefix tree of all article titles. 
This method surpassed some traditional pipelined approaches across various tasks based on Wikipedia.
Subsequent research efforts \cite{DSI,NCI,chen2022corpusbrain,seal,zhuang2022bridgingdsiqg} have continued to investigate and enhance the GR paradigm.
For example, 
\citet{zeng2023scalable} design a multi-stage training strategy to generalize GR from moderate-scale datasets \cite{naturalquestion} to large-scale datasets \cite{msmarco}.

\heading{Advantages}
The \ac{GR} paradigm offers several advantages:
\begin{enumerate*}[label=(\roman*)]
    \item It enables end-to-end optimization, allowing the model to be trained towards the global objective. This means that the entire retrieval process, including both document representation and ranking, can be optimized jointly.
    \item During inference, given a query, the generative model generates docids based on a small-sized vocabulary with beam search. This approach improves retrieval efficiency by eliminating the need for a heavy traditional index, where all documents in the corpus need to be matched against the query for dense retrieval methods.
\end{enumerate*}
% This paradigm shows promise in achieving more efficient and effective retrieval by leveraging the power of large-scale generative models and optimizing the retrieval process in end-to-end manner.

\heading{Limitations}
There are several limitations to the \ac{GR} paradigm.
For example, existing work optimizes the model with query-docid pairs by straightforward MLE, 
which only supports finding the most relevant docids. 
For queries with multiple relevant docids with multiple relevance grades, the relative order of these relevant docids in the ranked list is randomized,
resulting in sub-optimal overall relevance of the ranked list.
In this work, we optimize the ranked docid list in a listwise manner and calibrate the generation probabilities of docids within the ranked docid list generated by a beam search-based strategy. To the best of our knowledge, this work is the first attempt to perform listwise optimization in GR.

\section{Conclusion and future work}
\label{sec:conclusion}

In this paper, to better align with practical retrieval needs of generating a ranked list of results in response to a query, we propose to directly model ranked docid lists in \acl{GR}, so that docid lists instead of invididual docids are used as instances in learning. 
Inspired by position-aware ListMLE in LTR, and considering the characteristics of GR, we maximize the $i$-th conditional likelihood of a Plackett-Luce model given the top $i-1$ docids. 
Furthermore, to address the issue of beam search decoding in GR, we design relevance calibration to optimize the order of docids in the list. 
By conducting comprehensive experiments, we have substantiated that our approach exhibits superior effectiveness compared to existing GR methods. 

ListGR has several limitations that give rise to interesting lines of future work:
\begin{enumerate}[label=(\roman*)]
     \item This work represents our initial exploration of listwise GR, and there are many other listwise approaches \cite{10.1145/1273496.1273513listnet,xia2008listwiselistmle} in the LTR literature. In the future, we will continue to explore and optimize this work from multiple perspectives. For example, we will investigate how to design position weights in the loss function from a theoretical perspective to make it more suitable for specific use cases. Additionally, we may generate the entire list using a single beam instead of multiple beams, in order to alleviate the impact of beam search decoding on performance.
    
    \item  In this study, we did not extensively address the design of docids. It is worth noting that the choice of docids can significantly impact both the learning process and retrieval effectiveness. Similar to most existing GR approaches, we assumed that docids are unrelated to the retrieval model and did not optimize them. It is desirable to incorporate the generation and optimization of docids into the model optimization process, allowing for joint learning of docids that are well-suited for GR.
    
    \item The paper emphasizes modeling relevance at the list level but acknowledges that relevance should not be the sole focus \cite{shah2022situating}. LM-based search systems prioritize technology over user-centric aspects, necessitating further development in user interaction and personalization modules. Addressing bias and ensuring controllable and trustworthy search systems are also important topics, along with traceability and interpretability of retrieval structures.
\end{enumerate}

\section*{Reproducibility}
To facilitate reproducibility in this paper, we have only used open datasets. Detailed experimental results and settings are available at \url{https://github.com/lightningtyb/ListGR}.

\begin{acks}
This work was funded by the Strategic Priority Research Program of the CAS under Grants No. XDB0680102, 
the National Key Research and Development Program of China under Grants No. 2023YFA1011602,  
the project under Grants No. JCKY2022130C039, 
the Lenovo-CAS Joint Lab Youth Scientist Project, 
the CAS Project for Young Scientists in Basic Research under Grant No. YSBR-034, and the Innovation Projectof ICT CAS under Grants No. E261090.

This work was also (partially) funded by 
the Hybrid Intelligence Center, a 10-year program funded by the Dutch Ministry of Education, Culture and Science through the Netherlands Organisation for Scientific Research, \url{https://hybrid-intelligence-centre.nl}, 
project LESSEN with project number NWA.1389.20.183 of the research program NWA ORC 2020/21, which is (partly) financed by the Dutch Research Council (NWO),
and
the FINDHR (Fairness and Intersectional Non-Discrimination in Human Recommendation) project that received funding from the European Union’s Horizon Europe research and innovation program under grant agreement No 101070212.

We thank our anonymous reviewers for helpful and constructive feedback that helped us to improve the paper. All content represents the opinion of the authors, which is not necessarily shared or endorsed by their respective employers and/or sponsors.
\end{acks}

\bibliographystyle{ACM-Reference-Format}
\bibliography{references}

\end{document}